\documentclass[conference]{IEEEtran}
\usepackage{blindtext, graphicx}
\usepackage{listings}
\usepackage[numbers]{natbib}

\usepackage{array}
\usepackage{booktabs}
\usepackage{multirow}
\usepackage{siunitx}
\usepackage{color,soul}
\usepackage{caption}
\usepackage{subcaption}
\usepackage{array}
\usepackage{booktabs}
\usepackage{hyperref}
\usepackage{hyperref}
\usepackage[utf8]{inputenc}
\usepackage{mathtools, nccmath}

\usepackage{amsmath}
\usepackage{amssymb}
\usepackage[mathlines]{lineno}
\usepackage{lineno}

\usepackage{algorithm} 
\usepackage{algpseudocode}

\usepackage[figurename=Fig.]{caption}

\usepackage{array}
\usepackage{multirow}
\usepackage[symbol]{footmisc}

\title{seUNet-Trans: A Simple yet Effective UNet-Transformer Model for Medical Image Segmentation}

\author{\IEEEauthorblockN{Tan-Hanh Pham\textsuperscript{*}}
\IEEEauthorblockA{Department of Mechanical and\\
Civil Engineering, Florida Institute of\\
Technology, USA\\
Email: tpham2023@my.fit.edu}
\and
\IEEEauthorblockN{Xianqi Li}
\IEEEauthorblockA{Department of Mathematics and \\
Systems Engineering, Florida Institute of\\
Technology, USA \\
Email: xli@fit.edu}
\and
\IEEEauthorblockN{Kim-Doang Nguyen\textsuperscript{*}}
\IEEEauthorblockA{Department of Mechanical and\\
Civil Engineering, Florida Institute of\\
Technology, USA\\
Email: knguyen@fit.edu}}
\begin{document}

\maketitle
\thispagestyle{empty}
\pagestyle{empty}

\begin{abstract}

Automated medical image segmentation is becoming increasingly crucial to modern clinical practice, driven by the growing demand for precise diagnosis, the push towards personalized treatment plans, and the advancements in machine learning algorithms, especially the incorporation of deep learning methods. While convolutional neural networks (CNN) have been prevalent among these methods,  the remarkable potential of Transformer-based models for computer vision tasks is gaining more acknowledgment. To harness the advantages of both CNN-based and Transformer-based models, we propose a simple yet effective UNet-Transformer (seUNet-Trans) model for medical image segmentation. In our approach, the UNet model is designed as a feature extractor to generate multiple feature maps from the input images, then the maps are propagated into a bridge layer, which is introduced to sequentially connect the UNet and the Transformer. In this stage, we approach the pixel-level embedding technique without position embedding vectors, aiming to make the model more efficient. Moreover, we apply spatial-reduction attention in the Transformer to reduce the computational/memory overhead. By leveraging the UNet architecture and the self-attention mechanism, our model not only retains the preservation of both local and global context information but also is capable of capturing long-range dependencies between input elements. The proposed model is extensively experimented on five medical image segmentation datasets including polyp segmentation to demonstrate its efficacy. Comparison with several state-of-the-art segmentation models on these datasets shows the superior performance of our proposed seUNet-Trans network. 

Keywords: Polyps, Colonoscopy, Medical image analysis, Deep learning, Vision transformers.

\end{abstract}


\section{Introduction}
\label{sec.intro}
Medical image segmentation involves identifying and extracting meaningful information from complex medical images, playing a crucial step in many clinical applications including computer-aided diagnosis, image-guided surgery, and treatment planning \citep{declinically}, \citep{ouyang2020video}. To date, manual segmentation by trained experts such as radiologists or pathologists remains the gold standards for delineating anatomical structures and pathological abnormalities. However, this process is costly, labor-intensive, and often requires significant experience. Deep learning-based models, on the other hand, have demonstrated outstanding performance in the automatic segmentation of objects of interest. This is attributed to their capability to discern and comprehend complex patterns and features within medical images. As a result, there is a significant demand for deep learning-driven automated medical image segmentation in clinical practice.


As a prominent subset of various image segmentation models, convolutional neural networks (CNN) have proven to be highly effective and greatly promising in numerous medical image segmentation tasks \citep{hesamian2019deep}, \citep{isensee2019automated}, especially UNet \citep{ronneberger2015u}, a type of fully convolutional network \citep{long2015fully}, consisting of a symmetric encoder and decoder architecture with skip connections to pass features from the encoder path to the decoder path.  However, due to the lack of ability to capture the long-range dependencies and global context information in images, these architectures typically produce inferior performance, particularly for target information that exhibits significant differences among patients in texture, shape, and size. To address these shortcomings, current research suggests implementing self-attention mechanisms grounded in CNN attributes \citep{schlemper2019attention}, \citep{wang2018non}. It is worth noting that Transformer \citep{vaswani2017attention},  initially conceived for sequence-to-sequence tasks in natural language processing (NLP) frameworks and being emerged as alternative architectures that entirely abandon convolutional operators and relies exclusively on attention mechanisms \citep{vaswani2017attention},  has ignited significant debate within the computer vision (CV) community. In contrast to previous CNN-driven methods, Transformers not only excel at capturing global context information but also showcase enhanced adaptability for downstream tasks when pre-trained on a large scale. For example, the first fully self-attention-based vision transformers (ViTs) for image recognition was introduced in \citep{dosovitskiy2020image} and achieved competitive outcomes on ImageNet \citep{krizhevsky2012imagenet} using 2D image patches with positional embedding as an input sequence, provided it was pre-trained on an extensive external dataset. Detection transformer (DETR) \citep{carion2020end} employs a transformer-based approach as a fully end-to-end object detector, delving into the connections between objects and the overall image context for object detection. Segmentation Transformer (SETR) \citep{zheng2021rethinking} replaces the traditional encoders with transformers in the standard encoder-decoder networks, effectively attaining state-of-the-art (SOTA) outcomes in the task of natural image segmentation. While Transformer is good at capturing global context, it struggles to grasp fine-grained details, especially for medical images. To overcome this limitation, efforts have been made by researchers to integrate CNN- and Transformer-based models into each other. In particular, TransUNet \citep{chen2021transUNet} and TransFuse \citep{zhang2021transfuse} are the representative ones by combining the Transformer and UNet for medical image segmentation. 

As a continuous effort to harness the strengths of CNN and Transformer-based models, we introduce a novel UNet-Transformer model, named as seUNet-Trans, tailored for medical image segmentation. Within this framework, the UNet serves as a feature extractor, deriving multiple feature maps from the input images. These maps are then fed into a bridge layer, strategically placed to bridge the UNet and the Transformer components in a sequential manner. Notably, our approach employs a pixel-level embedding technique without position embedding vectors to enhance the model's efficiency. Furthermore, the Transformer head plays a central role in modeling the relationships and dependencies among input sequences, culminating in the generation of a prediction map for the input images. By leveraging the UNet architecture and the Transformer mechanism, our model not only retains the preservation of both local and global context information but also is capable of capturing long-range relationships between input elements.

The rest of this paper is organized as follows. Section II provides an overview of related work in the field of automated medical image segmentation. Section III presents the architecture of the proposed seUNet-Trans model. Section IV focuses on numerical experiments
and comparisons with other state-of-the-art segmentation models. Section IV draws the conclusion for our work. 

\section{Related work}
\label{sec.relatedwork}
In this section, we begin by providing an overview of the commonly used CNN-based methods for medical image segmentation. We then explore recent advancements in the application of transformers within the realm of computer vision, particularly in segmentation tasks. Finally, we highlight the standard techniques that merge both CNN and Transformer architectures.

\subsection{CNN-based Medical Image Segmentation}
Over the last decade, the field of medical image segmentation has witnessed remarkable achievements using CNNs, especially the FCN, UNet, and their variants. For instance, UNet++ \citep{zhou2018UNet++} introduces a set of nested and densely skip connections to minimize the discrepancy between the encoding and decoding process. Attention U-Net \citep{oktay2018attention} proposes an innovative attention gate method, which empowers the model to prioritize targets with varying sizes and exclude non-pertinent feature responses. Res-UNet \citep{diakogiannis2020resUNet} incorporates a weighted attention mechanism and a skip connection scheme \citep{he2016deep} to enhance the performance of retinal vessel segmentation. R2U-Net merges the advantages of residual networks with UNet to elevate its feature representation capabilities. The PraNet \citep{fan2020pranet}, a.k.a. the parallel reverse attention network, employs the parallel partial decoder (PPD) and reverse attention (RA) model for polyp segmentation. KiU-Net \citep{valanarasu2020kiu} designs a unique architecture that leverages both under-complete and over-complete features to improve the segmentation performance of small anatomical structures. DoubleU-Net \citep{jha2020doubleu} establishes a robust foundation for medical image segmentation by chaining two U-Nets and implementing atrous spatial pyramid pooling (ASPP). FANet \citep{tomar2022fanet}, during training, consolidates the mast from the previous epoch with the feature map of the current epoch. Given that these methods are anchored in CNNs, they inherently miss out on capturing long-range dependencies and understanding global contextual ties. 

\subsection{Transformer-based Medical Image Segmentation}
Transformers \citep{vaswani2017attention} were first developed for machine translations and have now achieved top-tier performance in various NLP tasks. Inspired by their successes, many efforts have been made to adapt Transformers for computer vision tasks. In particular, ViT \citep{dosovitskiy2020image} is the pioneering endeavor demonstrating that a solely transformer-based architecture can attain superior performance in image recognition, given pre-training on a substantial dataset. Utilizing ViT as an encoder, Segmenter \citep{strudel2021segmenter} provides a segmentation framework by proposing a mask transformer decoder to generate class embeddings. With a combination of a transformer-based hierarchical encoder and a lightweight multilayer perceptron (MLP), SegFormer \citep{xie2021segformer} offers a simple yet potent segmentation architecture. By integrating an additional control function into the self-attention module, MedT \citep{valanarasu2021medical} proposed a gated axial-attention that extends the existing transformer-based architecture. Swin Transformer \citep{liu2021swin} recently attracted great attention due to its exceptional performance on a number of benchmarks for tasks such as image classification, object detection, and semantic segmentation. In contrast to many previous transformer-based models, Swin Transformer proposes a hierarchical architecture whose representation is computed with shifted windows. This strategy enhances efficiency by restricting self-attention computation to non-overlapping local windows while also allowing for cross-window connection. The hierarchical structure combined with the shifted window technique as a backbone can benefit other network architecture. By incorporating Swin Transformer into the encoder and decoder of the U-shaped architecture, DS-TransUNet \citep{lin2022ds} proposes a novel deep medical image segmentation framework that can effectively capture the non-local dependencies and multiscale contexts for improving the semantic segmentation quality of varying medical images. Extensive numerical experiments across seven typical medical image segmentation tasks show the effectiveness of this framework.

\subsection{CNN-Transformer - based Medical Image Segmentation}
Despite transformer-based methods that can model the global context at all stages, they process inputs as 1D sequences which may result in low-resolution features, thereby lacking precise localization information. Simply resorting to direct upsampling to achieve full resolution doesn't effectively recover this information which therefore leads to an imprecise segmentation result. To address this issue, significant research efforts have been made to integrate CNN with the self-attention mechanism by characterizing global relationships of all pixels through the feature maps. TransUNet \citep{chen2021transUNet} is the first such framework combining Transformer with UNet and achieving SOTA performance on medical image segmentation tasks. TransFuse \citep{zhang2021transfuse}  proposes a shallow CNN-based encoder and transformer-based segmentation network in parallel to enhance the efficiency for modeling global contexts. Inspired by these works, we conduct further investigations. Specifically,  the UNet model is designed to extract and output multiple feature maps from the input images. Then these feature maps are fed into an introduced bridge layer, which plays the role of sequentially connecting UNet and Transformer, enhancing the practical performance of various medical image segmentation tasks significantly. 

\section{Methodology}
\label{sec.methodology}
\begin{figure*}
\begin{center}
\includegraphics[width= 0.995\textwidth]{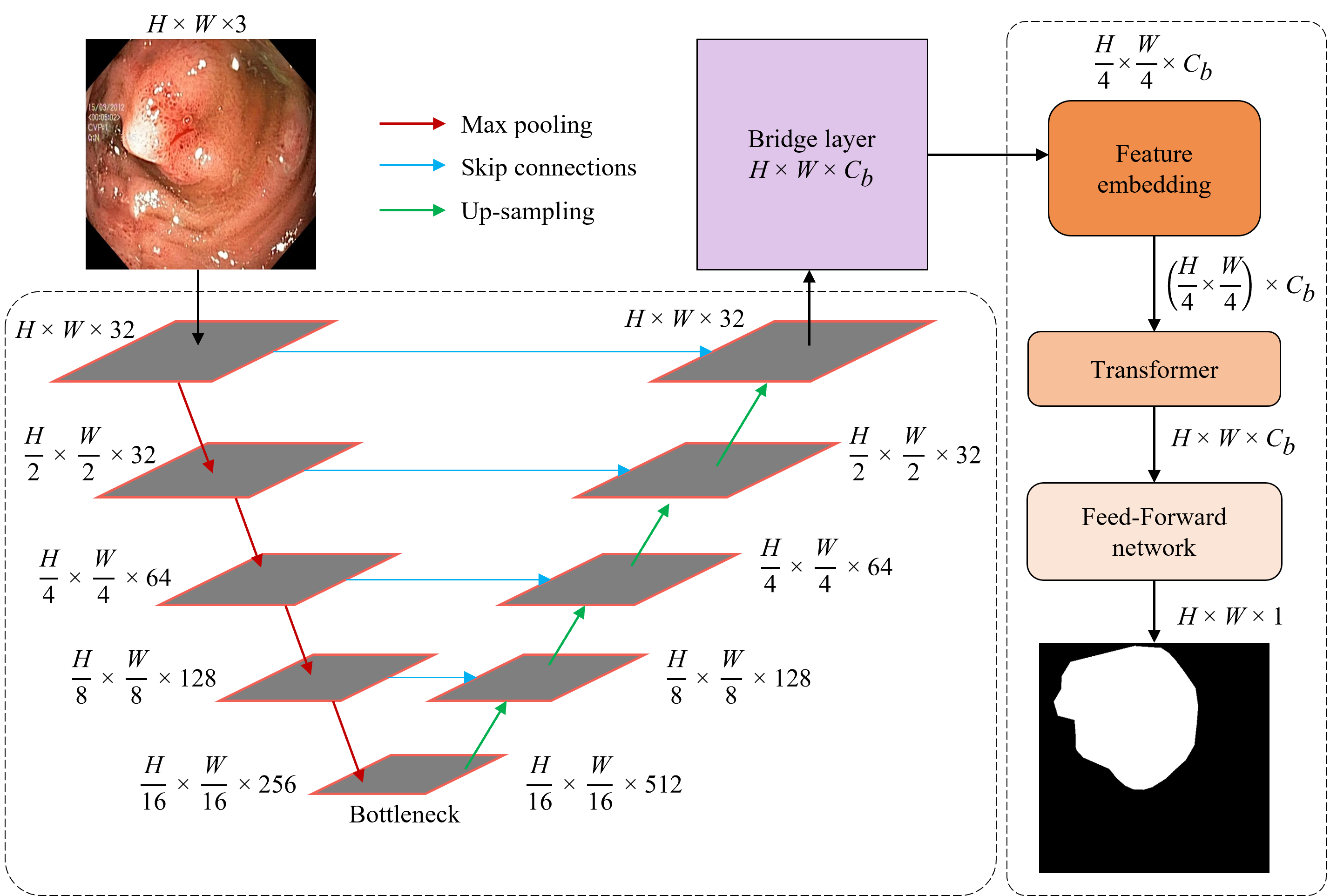}
\end{center}
\caption{The architecture of seUNet-Trans models.}
\label{fig.UNetTransformer} 
\end{figure*}

In this section, we introduce our proposed model in detail for medical image segmentation. Our model is comprised of a UNet as a backbone coupled with a Transformer head. The U-shaped backbone, consisting of an encoder and a decoder, processes input images to produce multiple feature maps, which are fed into specially designed bridge layers. Subsequently, the Transformer head processes the output from these bridge layers to yield the final prediction. The architecture of the seUNet-Trans model is shown in \ref{fig.UNetTransformer}.

\subsection{Encoder}
\label{subsec.encoder}
\begin{figure}
\begin{center}
\includegraphics[width= 0.25\textwidth]{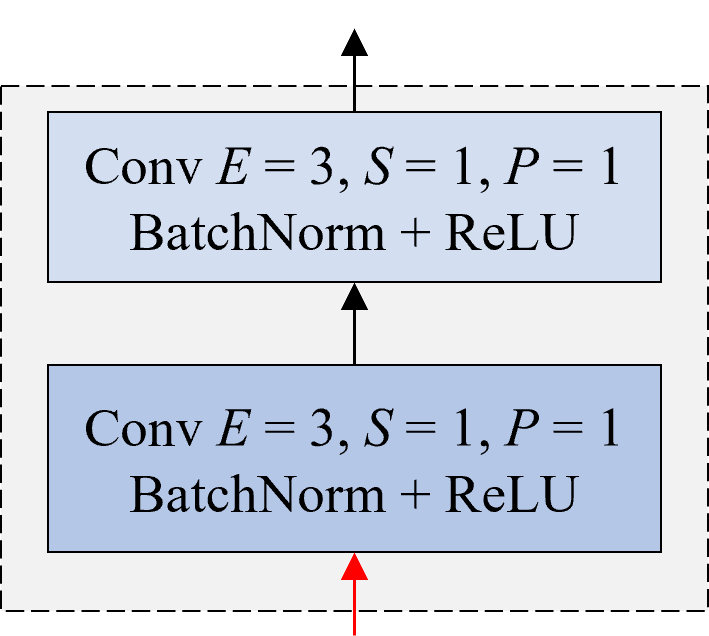}
\end{center}
\caption{UNet block.}
\label{fig.UNetblock} 
\end{figure}

The encoder's role is to extract features from the input images within the network. This is achieved through a series of convolutional layers, referred to as UNet blocks, succeeded by max-pooling layers. As the input images progress through these UNet blocks, their spatial dimensions are reduced, while the depth (or number of channels), of the feature maps increases. Based on \cite{ronneberger2015u}, we constructed the encoder section with four UNet blocks, with the specific design of a UNet block illustrated in Fig. \ref{fig.UNetblock}. 


The UNet block includes two convolutional neural networks (Conv) \citep{lecun1995convolutional} followed by a batch normalization function \citep{ioffe2015batch} and a rectified linear unit ReLU activation function \citep{nair2010rectified}. The structure of the UNet block can be formulated as:
\begin{flalign}\label{eq.UNetblock}
    \begin{aligned}
        \hat{F}_i & =   \text{ReLU}\left(\text{Batch}\left(\text{Conv}_{(C_{in}, C_h)}(F_{i-1})\right)\right), \\
        F_i & =   \text{ReLU}\left(\text{Batch}\left(\text{Conv}_{(C_h, C_o)}(\hat{F}_i)\right)\right), \forall{i} \geq 1.
    \end{aligned}
\end{flalign}
Where $\hat{F}_i$ and $F_i$ are intermediate and final feature maps for each UNet block, respectively. $C_{in}$, $C_h$, $C_o$ represent input, hidden and output layers, respectively. 

\subsection{Decoder}
\label{subsec.decoder}
The decoder component focuses on upsampling the encoded feature maps back to the input image size and mirrors the architecture of the encoder. Instead of max-pooling layers, it uses an up-convolution (or transpose convolution) to increase the spatial dimensions.
Skip connections, a critical component of UNet, are also employed to help the decoder retrieve spatial information lost during encoding. At each level in the decoder, the output from the corresponding encoder level (before pooling) is concatenated with the upsampled feature maps. After the skip connection, the concatenated features are passed through convolutional layers to refine the upscaled features. 



\begin{figure}
\begin{center}
\includegraphics[width= 0.3\textwidth]{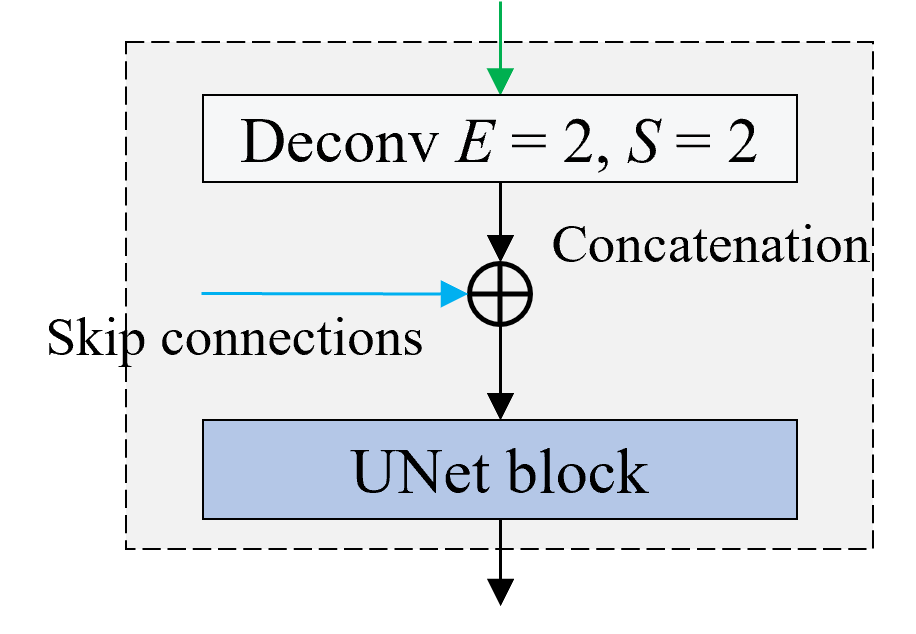}
\end{center}
\caption{Decoder block.}
\label{fig.Decoderblock} 
\end{figure} 

Distinct from the conventional decoder's final layer, which features a single channel for binary segmentation or several channels for multi-class segmentation with each channel representing the probability of a pixel belonging to a certain class, our enhanced layer outputs multiple feature maps, which will be fed into the next specially designed layers. 

\subsection {Bridge Layers}
\label{subsec.bridgelayer}

At the end of the decoder, the input images undergo a transforming process, yielding high-level features with identical dimensions but a different number of channels. To refine and expand these features, a convolution layer is employed with a kernel size of 1 and a batch size of 1. These layers act as the bridge between the UNet and the Transformer head, therefore denoted as ``bridge layers".

\subsection{Transformer head}
\label{subsec.transformerhead}
\begin{figure}
\begin{center}
\includegraphics[width= 0.25\textwidth]{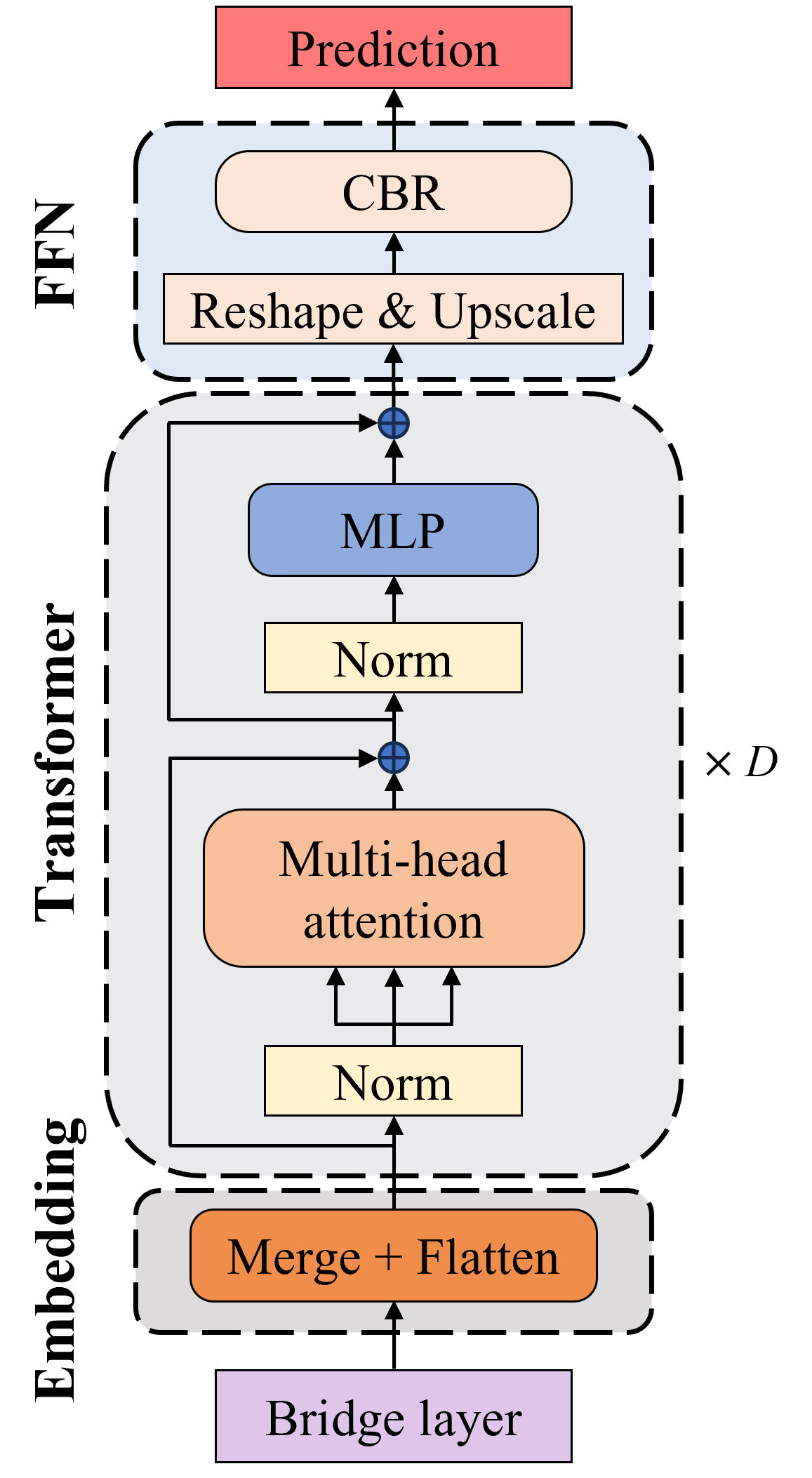}
\end{center}
\caption{Attention head in the seUNet-Trans model.}
\label{fig.AttentionHead} 
\end{figure}

The Transformer head begins by merging the features from the bridge layers. Subsequently, these merged features are flattened into sequences, and fed into the multi-head attention (MHA) mechanism. Then the output of the MHA is passed through the multi-layer perceptron (MLP), mainly used for mapping the input features to output features. Eventually, the output from the MLP is linearly upsampled, and processed by convolutional layers in the CBR block before outputting the final prediction. The structure of the Transformer head is shown in Fig. \ref{fig.AttentionHead}.

\subsubsection{Feature embedding}
\label{sec.subsubMergingEmbedding}
The bridge layers with the size of ($H$, $W$, $C_b$), height, width, and number of the bridge channels, are merged by using a convolutional layer with the kernel size $E$, stride $S$, and padding $P$ are 3, 4, and 1 respectively. After passing through the convolutional layer, the output resolution of the bridge layers is computed as:
\begin{flalign}\label{eq.shrinking}
    \begin{aligned}
        H_{out} & = \frac{(H - E + 2P)}{S} + 1, \\
        W_{out} & = \frac{(W - E + 2P)}{S} + 1. 
    \end{aligned}
\end{flalign}

In the context of image segmentation, our objective is to establish the relationship between pixels in the image. This can be accomplished through various methods, such as CNN-based techniques, attention mechanisms, and graph neural networks \citep{lecun1995convolutional, dosovitskiy2020image, pradhyumna2021graph}. In this particular study, we utilize the attention mechanism due to its effectiveness in capturing long-range features. We treat each pixel and its variations across different spatial dimensions (represented by various features in different channels) as a single input vector denoted as $a$. In other words, the merging features are flattened into sequences, and the dimensions of the sequences are $A \in \mathbb{R}^{N \times C_b}$, where $N = H_{out} \times W_{out}$.


Different from the Vision Transformer \citep{dosovitskiy2020image}, in this study, we opted not to use position embedding vectors during the input image flattening process. This choice is grounded in our approach of merging the input image and embedding the resultant features at the pixel level. Typically, the process of merging and embedding the bridge features into sequences can be formulated as: 
\begin{flalign}\label{eq.mergingFlattening}
    \begin{aligned}
        F_f & = \text{Flatten}\left(\text{Conv}_{(C_b, C_b)}\left(F_b\right)\right).
    \end{aligned}
\end{flalign}
Here, $F_b$ represents bridge layers, and $F_f$ is the embedding features.

\subsubsection{Transformer block}
\label{sec.subsubTransformerBlock}
The Transformer block consists of multi-head attention, multi-layer perceptron, LayerNorm, and residual connections, and it can be formulated as
\begin{flalign}
\label{eq.Transformerblock}
\begin{aligned}
\hat{F_i} &= \mbox{MHA}\left(\mbox{LN}\left(F_{i-1}\right)\right) + F_{i-1}, \\
F_i &= \mbox{MLP}(\mbox{LN}(\hat{F_i})) + \hat{F_i}.
\end{aligned}
\end{flalign}
Again, $\hat{F_i}$ and $F_i$ are intermediate and output layers of the $i^{th}$ Transformer block. For the first Transformer block or $i = 1$, the input is the embedding features ($F_f$).

In the MHA, the dependencies between sequences are computed by using cross-attention. In this step, the computational complexity is $N^2$ with $N$ as the number of input sequences. To reduce the computation, we used the sequence reduction technique implemented in \cite{wang2021pyramid} and \cite{xie2021segformer}, making it adaptable for high-resolution input images. Therefore, the complexity becomes $N^2/R$, where $R$ is the reduction rate.

The input sequences are divided into multiple heads $h$ in the MHA, in which the dimension of each head is $d_h$, $d_h = d_N/h$. In this study, we employ the length of the embedding vector $d_N = 64$, and the number of heads is $h = 4$. The attention in each head is calculated as 
\begin{equation}
\label{eq.selfattention}
\mbox{Attention}(Q, K, V) = \mbox{softmax}(\frac{QK^T}{\sqrt{d_h}})V,
\end{equation}
in which $Q \in \mathbb{R}^{N \times d_h}$, $K \in \mathbb{R}^{N/R \times d_h}$, and $V \in \mathbb{R}^{N/R \times d_h}$. 

Once the attention of each head is calculated, we combine all of them together to obtain the final attention matrix, 
\begin{flalign}
\label{eq.MHA}
\begin{aligned}
&\mbox{MultiHead}(Q, K, V) = \mbox{Concat}(\mbox{head}_1, ..., \mbox{head}_h)W^O, \\
&\mbox{and head}_e = \mbox{Attention}(QW_e^Q, KW_e^K, VW_e^V).
\end{aligned}
\end{flalign}
Where, $W_e^Q \in \mathbb{R}^{d_N \times d_h}$, $W_e^K \in \mathbb{R}^{d_N/R \times d_h}$, and $W_e^V \in \mathbb{R}^{d_N/R \times d_h}$ are parameter matrices over a head, and $W^O \in \mathbb{R}^{N \times d_N}$ is the total parameter matrix. 

The output from the MHA is added to its input through a residual connection. This connection facilitates the network's ability to learn residual information, representing the discrepancy between the expected output and the current estimate. Consequently, the network can adeptly capture and distribute gradient information during training, even in profoundly deep networks. Such a mechanism aids in efficiently training deeper neural architectures while addressing the vanishing gradient challenge.

Beyond the MHA, a Transformer block also encompasses a connected feed-forward network or MLP, which consists of two linear transformations with a GeLU activation \citep{hendrycks2016gaussian} between them. The aggregated features are first normalized before feeding into the MLP. Similar to the output from MHA, here we used another residual connection to add the MLP's output to its input. 
Equation \ref{eq.Transformerblock} describes the MHA and MLP procedure, in which the input features are mapped to the output features following the standard Transformer \citep{vaswani2017attention}. The process of the Transformer block can be repeated $D$ times, and in this study, we choose $D = 3$.

\subsubsection{Feed-Forward Network}
\label{sec.subsub.FFN}
The Feed-Forward Network (FFN) takes in the embedded sequences from the Transformer block to extract features and generate a prediction map. Given that the FFN operates on sequences as inputs, it becomes necessary to reshape these inputs to conform to the desired input shape ($H_{out}, W_{out}, C_b$). 

Furthermore, as computed in Section \ref{sec.subsubMergingEmbedding}, the input shape undergoes a merging operation, resulting in a fourfold reduction in size. Consequently, the reshaped features must be upsampled by a factor of four to match the original input shape ($H, W, C_b$). This upsampling process employs a bilinear interpolation function to increase the resolution of the feature maps. Mathematically, this step can be represented as follows:
\begin{flalign}
\label{eq.ReshapeUpscale}
\begin{aligned}
F_{rs} &= \mbox{Upscale}\left(\mbox{Reshape}\left(F_D\right)\right).
\end{aligned}
\end{flalign}
Here, $F_{rs}$ represents the upsampled feature maps after reshaping, and $F_D$ is the features from the Transformer block D, the final Transformer block.

After getting the upsampled features, they are fed into the CBR block for further processing, ultimately yielding the final prediction map. The CBR block, named for its convolutional layers, batch normalization, and ReLU activation, plays a vital role in feature refinement and spatial enhancement, enabling the network to capture intricate patterns and relationships within the data. The CBR consists of three convolutional layers, in which the first two layers with kernels size $E$ of $3\times3$ are followed by batch normalization and ReLU activation, while the third layer with a kernel size $E$ of $1\times1$ takes in features from previous layers and directly outputs the final prediction map $M$. Mathematically, this can be represented as follows:
\begin{flalign}\label{eq.CBR}
    \begin{aligned}
        \hat{F} & =   \text{ReLU}\left(\text{Batch}\left(\text{Conv}_{(C_b, C_{h1})}(F_{rs})\right)\right), \\
        \hat{F} & =   \text{ReLU}\left(\text{Batch}\left(\text{Conv}_{(C_{h1}, C_{h2})}(\hat{F})\right)\right),\\
        M & = \text{Conv}_{(C_{h2}, 1)}(\hat{F}).
    \end{aligned}
\end{flalign}
Again, $\hat{F}$ is the intermediate output of the CBR block. $C_{h1}$ and $C_{h2}$ are the hidden $1^{st}$ and $2^{nd}$ convolutional layers, respectively. In this study, we build the seUNet-Trans models for medical image segmentation. Hence, the final prediction $M$ is the binary image (one class). 

\section{Experiment and Evaluation} 
\label{sec.experiment}
In this section, we compare our proposed model with the state-of-the-art (SOTA) models in medical image segmentation using publicly available datasets. We first describe the datasets and outline the employed metrics to evaluate the model's efficacy. Further, specifics regarding the training and optimization processes are detailed in this section's conclusion.


\subsection{Dataset}
\label{subsec.dataset}
The seUNet-Trans models are trained on the Polyp Segmentation (Kvasir-SEG, CVC-ClinicDB, CVC-ColonDB, EndoScene), ISIC 2018, GlaS, and 2018 Data Science Bowl datasets, which are widely recognized and frequently used for evaluating various medical segmentation models such as DS-TransUNet, PraNet, and ColonSegNet \citep{lin2022ds, fan2020pranet, jha2021real}. 

\begin{table*}
    \centering
    \caption{Published datasets for training and testing the seUNet-Trans models.}
    \label{table.dataset}
    
    \begin{tabular}{lc|c|c|c}
        \toprule
        \multicolumn{2}{c|}{Dataset} & Size & {\begin{tabular}{@{}c@{}}Training \\ images\end{tabular}} & {\begin{tabular}{@{}c@{}}Test \\ images\end{tabular}}\\
        \midrule \addlinespace[-0.25ex] \midrule
        
        \multicolumn{2}{l|}{Kvasir-SEG} & $512\times512$ & 880 & 120 \\
        \midrule
        \multicolumn{2}{l|}{CVC-ClinicDB} & $384\times384$ &  550 & 62\\
        \midrule
        
        \multirow{4}{*}{\begin{tabular}{@{}lc@{}}Mixing Polyp \\segmentation \end{tabular}} & \multicolumn{1}{|l|}{Kvasir-SEG} & \multirow{4}{*}{$384\times384$} & 900 & 100 \\
        & \multicolumn{1}{|l|}{CVC-ClinicDB}&  &  550 & 62\\
        & \multicolumn{1}{|l|}{CVC-ColonDB}&  &  0 & 380\\
        & \multicolumn{1}{|l|}{EndoScene}& &  0 & 60\\
        \midrule
        
        \multicolumn{2}{l|}{GlaS}&  $128\times128$ & 85 & 80\\
        \midrule
        \multicolumn{2}{l|}{ISIC 2018} &  $256\times256$ & 2075 & 519\\
        \midrule
        \multicolumn{2}{l|}{2018 Data Science Bowl} & $256\times256$ & 536 & 134\\

        \bottomrule
    \end{tabular}

\end{table*}

For data preprocessing, we first standardized the dataset by resizing the images to a uniform scale. Subsequent to this, we divided the preprocessed dataset into separate training and test datasets. Table \ref{table.dataset} provides a comprehensive overview of the image divisions and the resized resolutions for various training scenarios. Notably, when dealing with the mixing Polyp segmentation case, we combined different datasets for training and testing tasks. In particular, the training set comprises 900 Kvasir-SEG images and 550 CVC-ClinicDB images, while the test set includes 100 Kvasir-SEG images, 62 CVC-ClinicDB images, 380 CVC-ColonDB images, and 60 EndoScene images following \citep{lin2022ds}.

\begin{figure}
\begin{center}
\includegraphics[width= 0.475\textwidth]{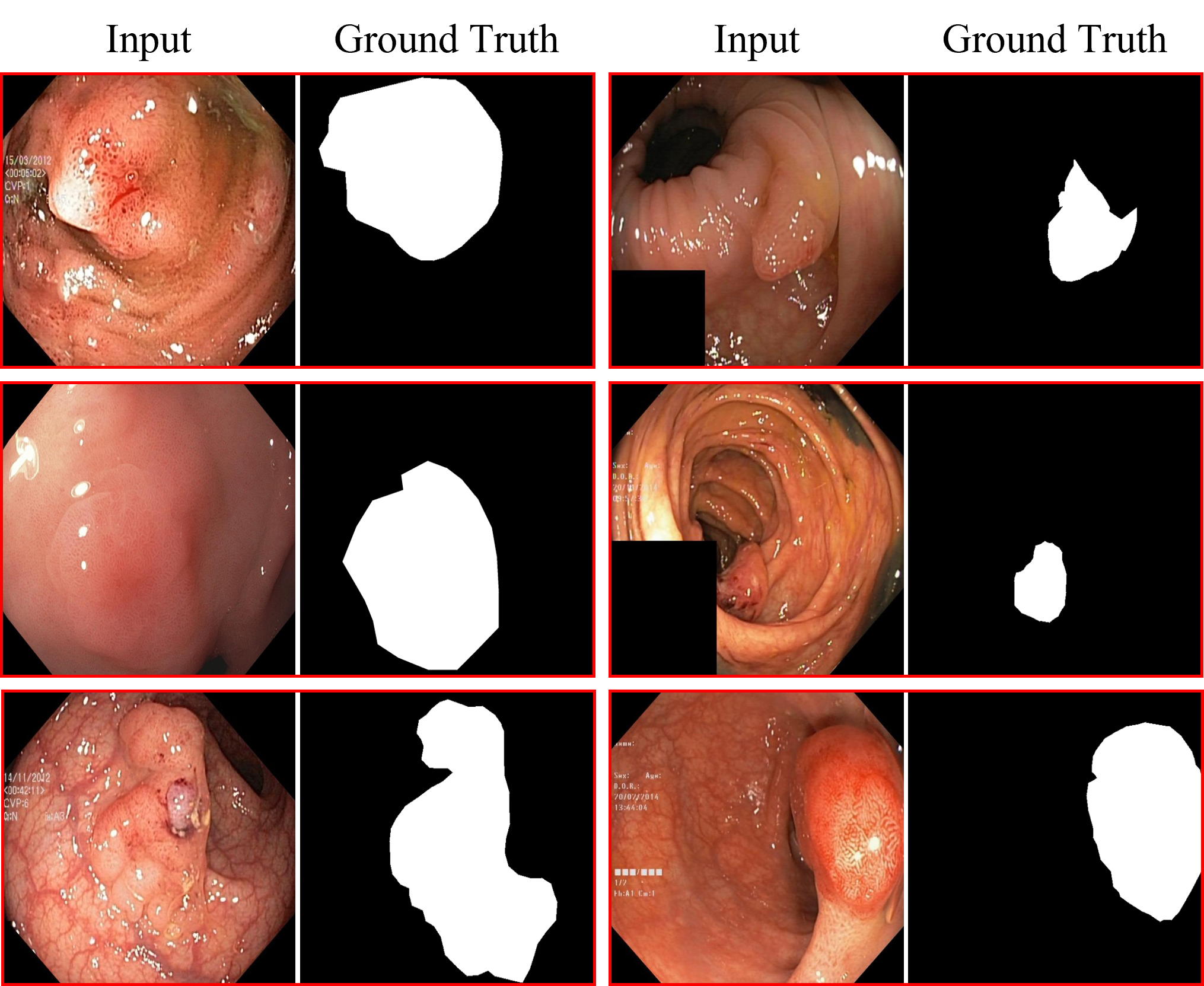}
\end{center}
\caption{Visual illustration of input images and their corresponding ground truths (segmentation maps) from the Kvasir-SEG dataset.}
\label{fig.dataset} 
\end{figure} 

In addition to the mixing Polyp segmentation, we have conducted training for our proposed seUNet-Trans models using the Kvasir-SEG or CVC-ClinicDB datasets separately. The Kvasir-SEG dataset is partitioned into 880 images for training and 120 for testing, and the CVC-ClinicDB dataset contains 550 images for training with a test set of 62 images. Visual representations from the Kvasir-SEG dataset are displayed in Figure \ref{fig.dataset}, where the input is presented as an RGB image, and the corresponding output is a binary segmentation mask. 
For the GlaS, ISIC 2018, and 2018 Data Science Bowl datasets, the training dataset contains 85, 2075, and 536 images, respectively, while the test dataset comprises 80, 519, and 134 images for each.

\subsection{Evaluation metrics}
To evaluate the performance of the seUNet-Trans models, we employ standard segmentation metrics including mean IoU (mIoU), mean Dice Coefficient (mDC) or mDC score, mean Precision (mPre.), and mean Recall (mRec.). These metrics are calculated by comparing the model's predictions against the ground truths across the entire dataset of T images, and are given as follows:

\begin{flalign}
\label{eq.metrics}
    \begin{aligned}
        \mbox{mIoU} &= \frac{1}{T}\sum_{t=1}^{T}\frac{TP_t}{TP_t+FP_t+FN_t}, \\
        \mbox{mDC} &= \frac{1}{T}\sum_{t=1}^{T}\frac{2TP_t}{2TP_t+FP_t+FN_t},\\
        \mbox{mPre.} &= \frac{1}{T}\sum_{t=1}^{T}\frac{TP_t}{TP_t+FP_t},\\
        \mbox{mRec.} &= \frac{1}{T}\sum_{t=1}^{T}\frac{TP_t}{TP_t+FN_t}.
    \end{aligned}
\end{flalign}

Where $TP, TN, FP, FN$ are the True Positive, True Negative, False Positive, and False Negative, respectively. 

\subsection{Model training}
Our seUNet-Trans models have been developed utilizing the PyTorch 1.13.1 deep learning framework. For the training process, we employ the 'AI-Panther' high-performance computing infrastructure, furnished with A100 SXM4 GPUs, hosted at the Florida Institute of Technology.

As mentioned in section \ref{sec.subsub.FFN}, the final prediction is the binary image. This prompts us to use the binary cross-entropy (BCE) loss as the objective function during the training. The BCE loss measures the difference between predicted and ground truth images. Each pixel in the prediction, $M_x$, with values ranging from 0 to 1, is compared to its corresponding pixel in the ground truth, $Y_x$. Consequently, the average loss function for a pair of prediction and ground truth images is formulated as follows:

\begin{flalign}
\label{eq.lossfunction}
\begin{aligned}
\text{Avg. BCE} (\theta) =& -\frac{1}{X} \sum_{x=1}^{X} \Big[ Y_x \log\big(M_x(\theta)\big) \\ & + (1 - Y_x) \log\big(1 - M_x(\theta)\big) \Big].
\end{aligned}
\end{flalign}
Again, $M$ is the prediction, $Y$ is the ground truth, and $X$ is the total number of pixels in the prediction or ground truth.

We employ the Adam optimizer \citep{kingma2014adam} with a learning rate of 0.0001, weight decay of 0.0001, and a batch size of 8 for estimating model parameters $\theta$. Checkpoints are saved every 10 epochs and are subsequently loaded for evaluating the model on the test dataset upon the training phase's completion.

\section{Experimental Results} 
\label{sec.resultanddiscussion}
In this section, we detail the performance of our model across all utilized datasets, providing a thorough comparative analysis with state-of-the-art (SOTA) models. We present our findings in tabular form and include SOTA results referenced from \cite{lin2022ds} to facilitate a clear and comprehensive comparison. For a visual illustration, Figures \ref{fig.Kvasir-SEG_and_ClinicDB} to \ref{fig.2018Bowl} display predicted results by seUNet-Trans models for a selection of representative images, along with those produced by SOTA models. To quantitatively evaluate our model's performance, Table \ref{table.ResultKvasirdataset} to Table \ref{table.2018Bowl} present the calculated metric scores obtained across the various datasets.

As mentioned in Section \ref{sec.subsubTransformerBlock}, we implemented a sequence reduction technique to mitigate the computational complexity in the attention head. Specifically, we adopted reduction ratios $R$ with values of 4, 2, and 1, which correspond to strides $S$ of 2, 4, and 8, respectively. It's worth noting that a smaller stride retains more information but increases the number of input sequences, which can lead to longer training times. With the pairs of combinations of reduction ratios and strides, we developed three distinct models: seUNet-Trans-L, seUNet-Trans-M, and seUNet-Trans-S. Therefore, in this section, we compare all our models' performance with other models.

\subsection{Results on Kvasir-SEG}
\label{sec.Kvasir-SEGresult}
\begin{figure*}
\begin{center}
\includegraphics[width= 1\textwidth]{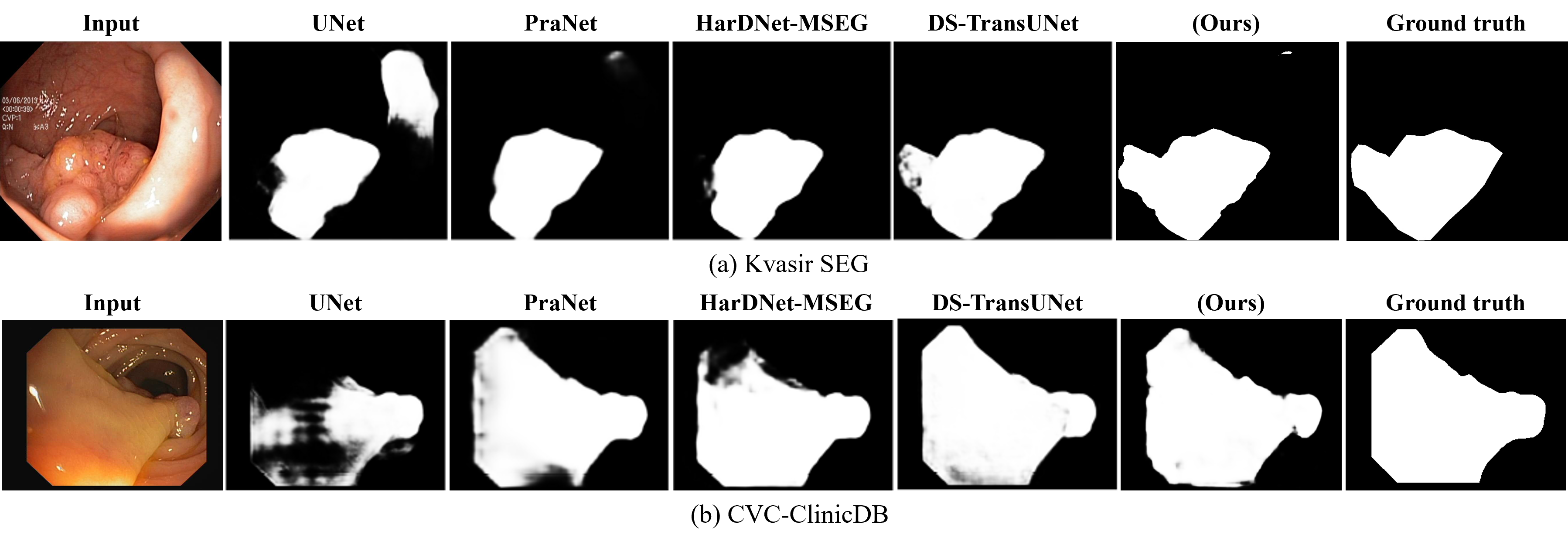}
\end{center}
\caption{Visualization of predictions of seUNet-Trans-M and the SOTA on the Kvasir-SEG and ClinicDB datasets. These images are partially taken from \cite{lin2022ds} for comparison purposes.}
\label{fig.Kvasir-SEG_and_ClinicDB} 
\end{figure*}

Figure \ref{fig.Kvasir-SEG_and_ClinicDB}a presents results predicted by the seUNet-Trans-M on the Kvasir-SEG dataset. We then perform a comparative analysis of these predictions against those produced by other models. The predictions of our models are comparable to those of other models, closely aligning with the objects in the ground truth. Furthermore, the calculated metric values are summarized in Table \ref{table.ResultKvasirdataset}, revealing that the seUNet-Trans models achieve impressive values of 0.919 for mDC, 0.850 for mIoU, 0.912 for mRec., and 0.926 for mPre.. Notably, seUNet-Trans outperforms other models in terms of mDC, and mPre., demonstrating superior performance. However, it's worth mentioning that the mIoU and mRec. scores of our models are relatively smaller than those of PraNet, HarDNet-MSEG, and DS-TransUNet. In addition, seUNet-Trans-M demonstrates consistent performance among our models.

\begin{table}
    \centering
    \caption{Quantitative results of evaluation metrics for seUNet-Trans in comparison to SOTA models on the Kvasir-SEG.}
    \label{table.ResultKvasirdataset}
    
    \begin{tabular}{l|cccc}
        \toprule
        Methodology & mDC & mIoU & mRec. & mPre.\\
        \midrule \addlinespace[-0.25ex] \midrule
        U-Net \cite{ronneberger2015u} & 0.597&  0.471 & 0.617 & 0.672\\
        Res-UNet  \cite{xiao2018weighted}&  0.690  & 0.572  & 0.725 &  0.745\\
        ResUNet++  \cite{jha2019resUNet++}& 0.714  & 0.613  & 0.742  & 0.784\\
        DoubleU-Net \cite{jha2020doubleu} & 0.813  & 0.733  & 0.840  & 0.861\\
        FCN8 \cite{long2015fully} &  0.831 &  0.737  & 0.835  & 0.882\\
        PSPNet \cite{zhao2017pyramid} & 0.841 &  0.744 &  0.836  & 0.890\\
        HRNet  \cite{wang2020deep} & 0.845  & 0.759  & 0.859 &  0.878\\
        DeepLabv3+ \cite{chen2018encoder} &  0.864 &  0.786 &  0.859  & 0.906\\
        FANet \cite{tomar2022fanet} & 0.880  & 0.810  & 0.906  & 0.901\\
        HarDNet-MSEG  \cite{huang2021hardnet}  & 0.904  & 0.848 &  0.923  & 0.907\\
        DS-TransUNet-L \cite{lin2022ds} & 0.913 &  \textbf{0.859} &  \textbf{0.936}  & 0.916\\
        \midrule
        seUNet-Trans-L (ours) & 0.914 &  0.841 &  0.901  & \textbf{0.927}\\

        seUNet-Trans-M (ours) & \textbf{0.919} &  0.850 &  0.912  & 0.926\\

        seUNet-Trans-S (ours) & 0.904 &  0.825 &  0.896  & 0.912\\
        
        \bottomrule
    \end{tabular}

\end{table}

\subsection{Results on CVC-ClinicDB}
Figure \ref{fig.Kvasir-SEG_and_ClinicDB}b shows the predicted results by seUNet-Trans-M and other models. As observed, the predictions by seUNet-Trans surpass not only those of the standard UNet model but also outperform SOTA models. A detailed comparison of the results is tabulated in Table \ref{table.Result_CVC-ClinicDB}, where the seUNet-Trans-M achieves remarkable performance, including a mDC of 0.945, mIoU of 0.895, mPre. of 0.951, and mRec. of 0.950. In contrast, the standard UNet model yields lower metric values with mDC, mIoU, mPre., and mRec. values of 0.872, 0.804, 0.868, and 0.917, respectively. This comparison underscores the superior performance of the seUNet-Trans on the CVC-ClinicDB dataset when compared to the baseline UNet model and other SOTA models.

\begin{table}
    \centering
    \caption{Quantitative results of evaluation metrics for seUNet-Trans in comparison to SOTA models on the CVC-ClinicDB.}
    \label{table.Result_CVC-ClinicDB}
    
    \begin{tabular}{l|cccc}
        \toprule
        Methodology & mDC & mIoU & mRec. & mPre.\\
        \midrule \addlinespace[-0.25ex] \midrule
        SFA \cite{fang2019selective}& 0.700 & 0.607&  - & -\\
        ResUNet-mod \cite{zhang2018road} &  0.779 & 0.455 & 0.668 & 0.888\\
        UNet++  \cite{zhou2018UNet++} & 0.794 & 0.729 & - & -\\
        U-Net \cite{ronneberger2015u} &  0.872 & 0.804&  0.868 & 0.917\\
        PraNet \cite{fan2020pranet} &  0.899 & 0.849&  - & -\\
        DoubleU-Net \cite{jha2020doubleu}&  0.924 & 0.861&  0.846 & \textbf{0.959}\\
        FANet  \cite{tomar2022fanet} & 0.936 & 0.894 & 0.934 & 0.940\\
        DS-TransUNet-L \cite{lin2022ds}& 0.942&  0.894&  0.950 & 0.937\\
        \midrule
        seUNet-Trans-L (ours) & 0.936 & 0.879 & 0.941 &  0.933\\
        
        seUNet-Trans-M (ours) & \textbf{0.945} &  \textbf{0.895} &  \textbf{0.951}  & 0.950\\

        seUNet-Trans-S (ours) & 0.938 & 0.888 & 0.936 &  0.945\\
        
        \bottomrule
    \end{tabular}

\end{table}

\subsection{Results on GlaS}
\begin{figure*}
\begin{center}
\includegraphics[width= 1\textwidth]{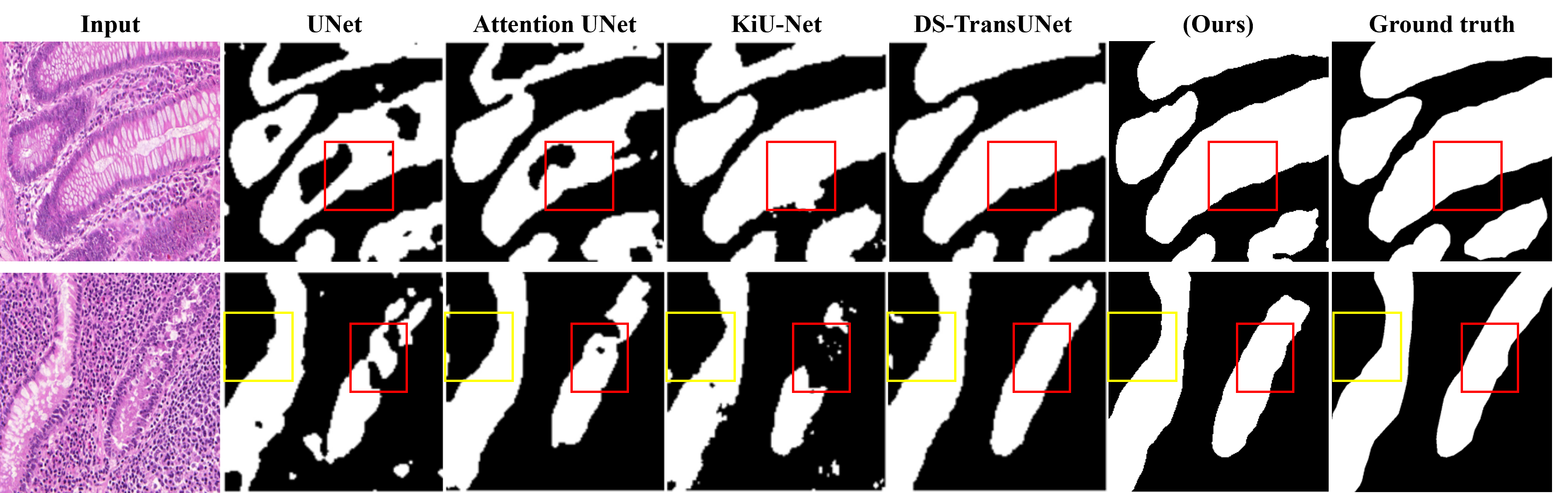}
\end{center}
\caption{Visualization of predictions of seUNet-Trans-M and the SOTA on the GlaS dataset. In these images, accurate regions are highlighted with red rectangles, while outliers among the models are indicated with yellow rectangles. These images are partially taken from \cite{lin2022ds} for comparison purposes.}
\label{fig.Result_GlaS} 
\end{figure*}

Figure \ref{fig.Result_GlaS} displays the predictions by the seUNet-Trans-M on the GlaS dataset. Compared to other SOTA models, seUNet-Trans-M stands out for its robust performance in gland segmentation. In addition, the results show that it not only outperforms most of the other models but also demonstrates comparability with DS-TranUNet. In terms of visualization, our model's predictions exhibit significantly higher accuracy, as indicated by the red rectangles, and demonstrate fewer outliers, highlighted by the yellow rectangles.

As mentioned in section \ref{subsec.dataset}, although the number of training samples in this dataset is very limited, our models perform well compared to other models. Table \ref{table.Result_GlaS} further reinforces seUNet-Trans's proficiency, revealing that its metric values surpass those of other models. Specifically, the seUNet-Trans-M achieves mDC and mIoU scores of 0.899 and 0.823, respectively, indicating its proficiency in gland segmentation on the GlaS dataset. In addition, all three variants of our models consistently yield superior results when compared to other models.

\begin{table}
    \centering
    \caption{Quantitative results of evaluation metrics for seUNet-Trans in comparison to SOTA models on the GlaS.}
    \label{table.Result_GlaS}
    
    \begin{tabular}{l|cccc}
        \toprule
        Methodology & mDC & mIoU & mRec. & mPre.\\
        \midrule \addlinespace[-0.25ex] \midrule

        Seg-Net \cite{badrinarayanan2017segnet}&0.786 &0.660 & - & -\\
        U-Net \cite{ronneberger2015u} & 0.796 & 0.672 & 0.845 & 0.778\\
        MedT \cite{valanarasu2021medical} &0.81 &0.696 &  -& -\\
        UNet++  \cite{zhou2018UNet++}  & 0.813 &0.696 & 0.857 & 0.798\\
        Attention UNet \cite{oktay1804attention} & 0.816 & 0.701 & 0.844 & 0.813\\
        KiU-Net \cite{valanarasu2020kiu} & 0.833 &0.728 & \textbf{0.889} & 0.809\\
        DS-TransUNet-L \cite{lin2022ds} & 0.878 &0.791 & 0.888 & 0.878 \\
        \midrule
        seUNet-Trans-L (ours) & 0.890 & 0.810 & 0.868 &  \textbf{0.923}\\

        seUNet-Trans-M (ours) & \textbf{0.899} & \textbf{0.823} & 0.886 &  0.920\\

        seUNet-Trans-S (ours) & 0.881 & 0.795 & 0.873 &  0.900\\

        \bottomrule
    \end{tabular}

\end{table}

\subsection{Results on ISIC 2018}
\begin{figure*}
\begin{center}
\includegraphics[width= 1\textwidth]{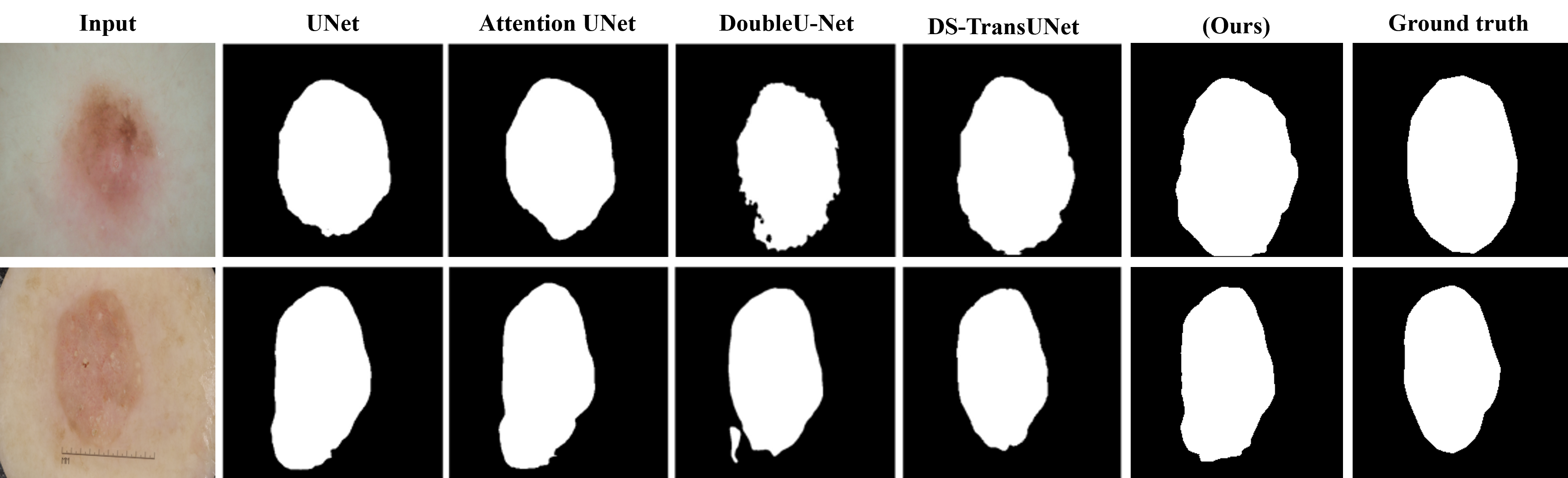}
\end{center}
\caption{Visualization of predictions of seUNet-Trans-M and the SOTA on the ISIC2018 dataset. These images are partially taken from \cite{lin2022ds} for comparison purposes.}
\label{fig.ISIC2018} 
\end{figure*}

Figure \ref{fig.ISIC2018} presents the predictions by our model on the ISIC 2018 dataset, and the corresponding metric values are detailed in Table \ref{table.ISIC2018}. In comparison to SOTA models, our models demonstrate strong performance on representative images.

In terms of the metric values, the seUNet-Trans-M still achieves commendable scores on ISIC 2018, with mDC, mIoU, mRec., and mPre. standing at 0.922, 0.854, 0.903, and 0.941, respectively. These metrics demonstrate the model's strong performance, compared to the SOTA models. 

\begin{table}
    \centering
    \caption{Quantitative results of evaluation metrics for seUNet-Trans in comparison to SOTA models on the ISIC2018.}
    \label{table.ISIC2018}
    
    \begin{tabular}{l|cccc}
        \toprule
        Methodology & mDC & mIoU & mRec. & mPre.\\
        \midrule \addlinespace[-0.25ex] \midrule
        U-Net \cite{ronneberger2015u} & 0.674&  0.549 & 0.708 & -\\
        Attention U-Net \cite{oktay1804attention} & 0.665 & 0.566&  0.717 & -\\
        R2U-Net \cite{alom2018recurrent} & 0.679&  0.581&  0.792 & -\\
        Attention R2U-Net \cite{alom2018recurrent} & 0.691 & 0.592&  0.726&  -\\
        BCDU-Net (d=3) \cite{azad2019bi} & 0.851&  - & 0.785 &  -\\
        FANet \cite{tomar2022fanet} & 0.8731 & 0.802 & 0.865 & 0.924\\
        DoubleU-Net  \cite{jha2020doubleu} & 0.896 & 0.821 & 0.878 & \textbf{0.946}\\
        DS-TransUNet-L \cite{lin2022ds}& 0.913 & 0.852 & \textbf{0.922} & 0.927\\
        \midrule
        seUNet-Trans-L (ours) & 0.918 &  0.849 &  0.900 & 0.938\\

        seUNet-Trans-M (ours) & \textbf{0.922} &  \textbf{0.854} &  0.903 & 0.941\\

        seUNet-Trans-S (ours) & 0.921 &  \textbf{0.854} &  0.906 & 0.937\\
        
        \bottomrule
    \end{tabular}

\end{table}

\subsection{Results on 2018 Data Science Bowl}
\begin{figure*}
\begin{center}
\includegraphics[width= 1\textwidth]{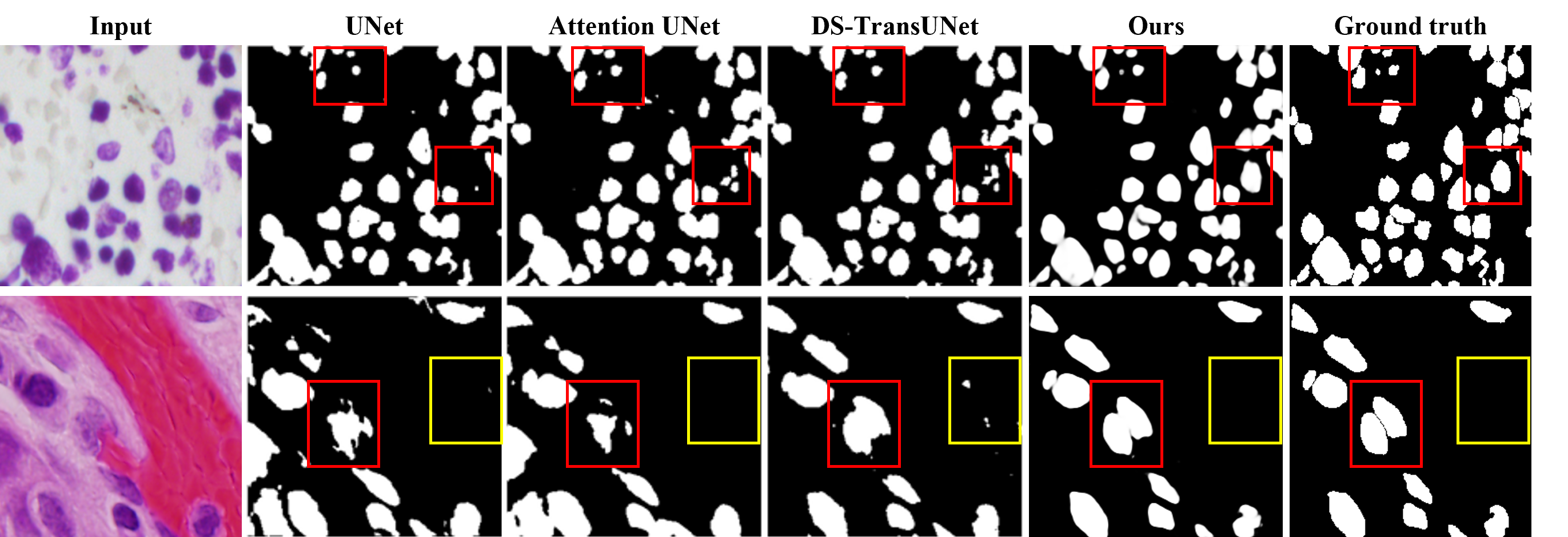}
\end{center}
\caption{Visualization of predictions of seUNet-Trans-M and the SOTA on the 2018 Data Science Bowl dataset. In these images, accurate regions are highlighted with red rectangles, while outliers among the models are indicated with yellow rectangles. These images are partially taken from \cite{lin2022ds} for comparison purposes.}
\label{fig.2018Bowl} 
\end{figure*}

The results of our seUNet-Trans-M on the 2018 Data Science Bowl dataset are visualized in Figure \ref{fig.2018Bowl}. Similar to Fig.\ref{fig.Result_GlaS}, we have compared the predicted masks generated by our models with those of others. In this context, the red rectangles draw attention to precision, while the yellow rectangles highlight outliers. As illustrated, our model's predictions do not include outliers and are more accurate compared to others.

Table \ref{table.2018Bowl} summarizes quantitative metrics among models. Once again, seUNet-Trans-M demonstrates its stability across the three variants and its superiority over SOTA models. Typically, the mDC and mIoU scores of seUNet-Trans-M are 0.928 and 0.867, relatively higher than other models. In comparison, DS-TransUNet, exhibits slightly lower metric values, with metric values of 0.922 and 0.861, respectively. This underscores the seUNet-Trans's proficiency on the 2018 Data Science Bowl dataset, with its predictions being notably free of outliers.

\begin{table}
    \centering
    \caption{Quantitative results of evaluation metrics for seUNet-Trans in comparison to SOTA models on the 2018 Data Science Bowl.}
    \label{table.2018Bowl}
    
    \begin{tabular}{l|cccc}
        \toprule
        Methodology & mDC & mIoU & mRec. & mPre.\\
        \midrule \addlinespace[-0.25ex] \midrule
        U-Net \cite{ronneberger2015u} & 0.757 & 0.910 & - & -\\
        UNet++ \cite{zhou2018UNet++} & 0.897 & 0.926 & - & -\\
        Attention UNet \cite{oktay1804attention} & 0.908 & 0.910 & - & 0.916\\
        DoubleU-Net  \cite{jha2020doubleu} & 0.913 & 0.841 & 0.641 & \textbf{0.950}\\
        FANet \cite{tomar2022fanet} & 0.918 & 0.857 & 0.922 & 0.919\\
        DS-TransUNet-L \cite{lin2022ds}& 0.922 & 0.861 & \textbf{0.938} & 0.912\\
        
        \midrule
        seUNet-Trans-L (ours) & 0.926 &  0.862 &  0.894 & 0.960\\

        seUNet-Trans-M (ours) & \textbf{0.928} &  \textbf{0.867} &  0.911 & 0.947\\

        seUNet-Trans-S (ours) & 0.914 &  0.842 &  0.884 & \textbf{0.950}\\
        
        \bottomrule
    \end{tabular}

\end{table}

\subsection{Results on mixed Polyp segmentation}
\label{sec.mixresult}
When training and testing with distinct datasets, it becomes evident that seUNet-Trans-M consistently yields reliable results. Therefore, for this experiment, we have employed this model to train and evaluate the mixed dataset.

\begin{figure*}
\begin{center}
\includegraphics[width= 1\textwidth]{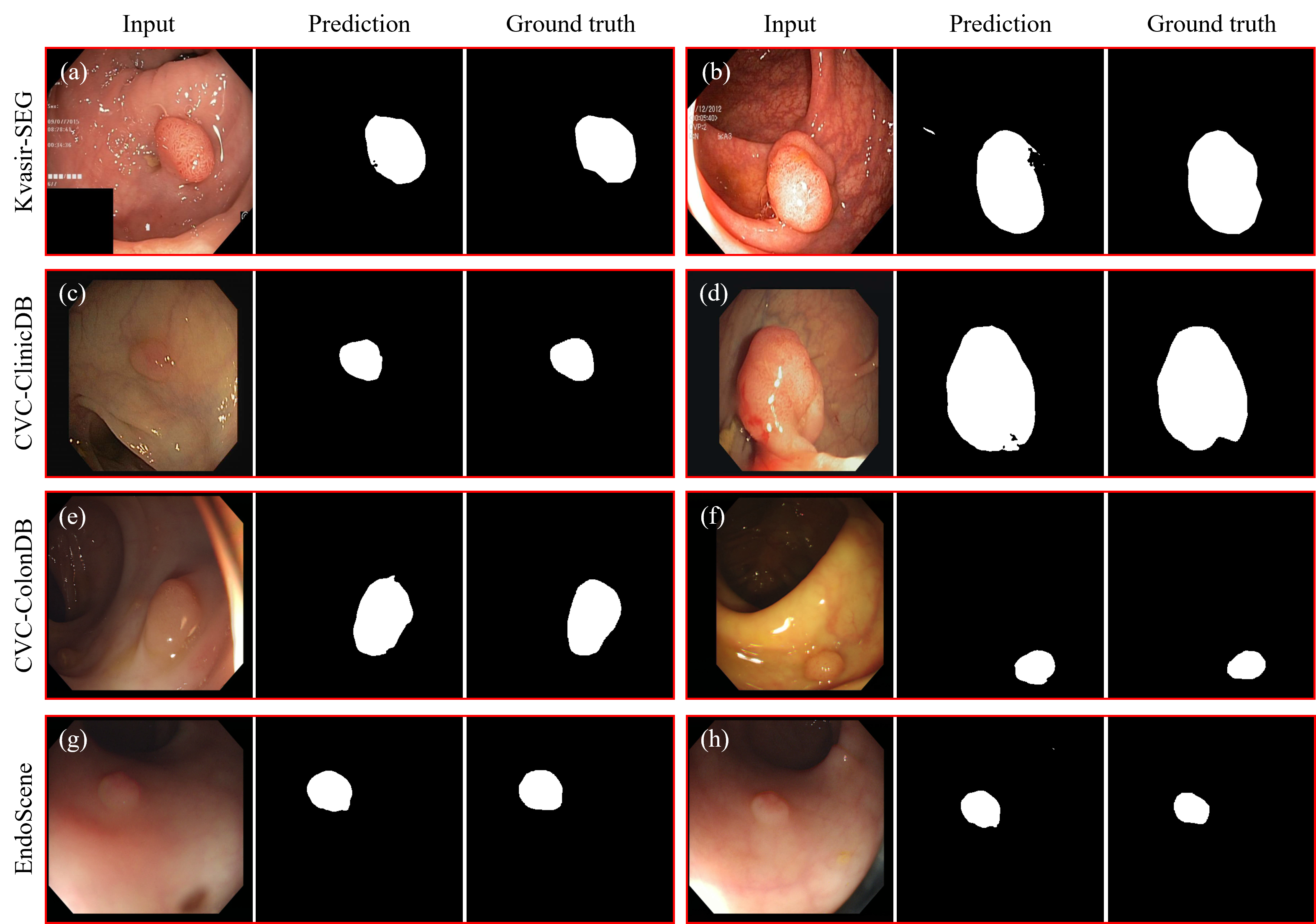}
\end{center}
\caption{Visualization of predictions produced by seUNet-Trans-M in the context of the mixing Polyp Segmentation experiment. Panels (a) and (b) pertain to CVC-ClinicDB, panels (c) and (d) to Kvasir-SEG, panels (e) and (f) to CVC-ColonDB, and panels (g) and (h) to EndoScene representations.}
\label{fig.mix} 
\end{figure*}

As described in section \ref{subsec.dataset}, seUNet-Trans-M was trained on a combined dataset comprising four distinct datasets for the mixed Polyp segmentation case. Figure \ref{fig.mix} illustrates our model's performance on the test set, in which its mDC and mIoU surpass those of SOTA models such as U-Net, PraNet, and DS-TransUNet.

Specifically, seUNet-Trans-M attains impressive mDC and mIoU scores of 0.942 and 0.913, respectively, on the Kvasir dataset, and 0.945 and 0.915 on the ClinicDB dataset, as highlighted in Table \ref{table.mix}. Remarkably, even on datasets it wasn't explicitly trained on, including ColonDB and EndoScene, the seUNet-Trans demonstrates exceptional predictive accuracy. On ColonDB, it attains mDC and mIoU of 0.905 and 0.864, respectively, while on EndoScene, these metrics stand at 0.903 and 0.861, showcasing the model's robustness and adaptability. 

\begin{table*}
    \centering
    \caption{Quantitative results of evaluation metrics for seUNet-Trans in comparison to SOTA models across four different datasets.}
    \label{table.mix}

    \begin{tabular}{lcc|cc|cc|cc|cc}
    \toprule
    \multirow{2}{*}{Methodology} & \multicolumn{2}{c}{Kvasir} & \multicolumn{2}{c}{ClinicDB} & \multicolumn{2}{c}{ColonDB}  & \multicolumn{2}{c|}{EndoScene} & \multicolumn{2}{c}{Average}\\ \addlinespace[0.5ex]
    \cline{2-11}  \\[-1em]
    & mDC & mIoU & mDC & mIoU & mDC & mIoU & mDC & mIoU & mDC & mIoU\\

    \midrule \addlinespace[-0.25ex] \midrule
    U-Net \cite{ronneberger2015u} & 0.818 &  0.746  & 0.823 &  0.755 &  0.512  & 0.444 &  0.398  & 0.335  &  0.652 &  0.581\\
    U-Net++ \cite{zhou2018UNet++}  & 0.821 &  0.743 &  0.794 &0.729 & 0.483 &  0.410 &  0.401 &  0.344  &  0.641  & 0.570\\
    PraNet \cite{fan2020pranet} &  0.898  & 0.840 &  0.899 &  0.849  & 0.709  & 0.640  & 0.871  & 0.797  & 0.800  & 0.739\\
    HarDNet-MSEG \cite{huang2021hardnet}&  0.912  & 0.857  & 0.932  & 0.882  & 0.731 &  0.660 &  0.887  & 0.821 &  0.828 &  0.767\\
    TransFuse-L  \cite{zhang2021transfuse} &  0.918  & 0.868  & 0.934  & 0.886 &  0.744  & 0.676 &  0.904  & 0.838   & 0.847 &  0.786\\
    DS-TransUNet-L \cite{lin2022ds} & 0.935  & 0.889 & 0.936  & 0.887 &  0.798  & 0.722 &  \textbf{0.911}  & 0.846  & 0.868  & 0.806\\
    \midrule
    seUNet-Trans-M (ours) & \textbf{0.942} & \textbf{0.913} & \textbf{0.945} & \textbf{0.915} & \textbf{0.905} & \textbf{0.864} & 0.903 & \textbf{0.861} & \textbf{0.934} & \textbf{0.899} \\

    \bottomrule
    \end{tabular}

\end{table*}

\subsection{Results on `mislabeling' data}



We conducted a comprehensive comparison of our models across a range of specific provided images, spanning from Section \ref{sec.Kvasir-SEGresult} to Section \ref{sec.mixresult}. In this section, we aim to present more intuitive results that highlight a clearer representation of the capabilities of our seUNet-Trans models.

\begin{figure*}
  \begin{subfigure}{\textwidth}
    \centering
    \includegraphics[width=0.95\linewidth]{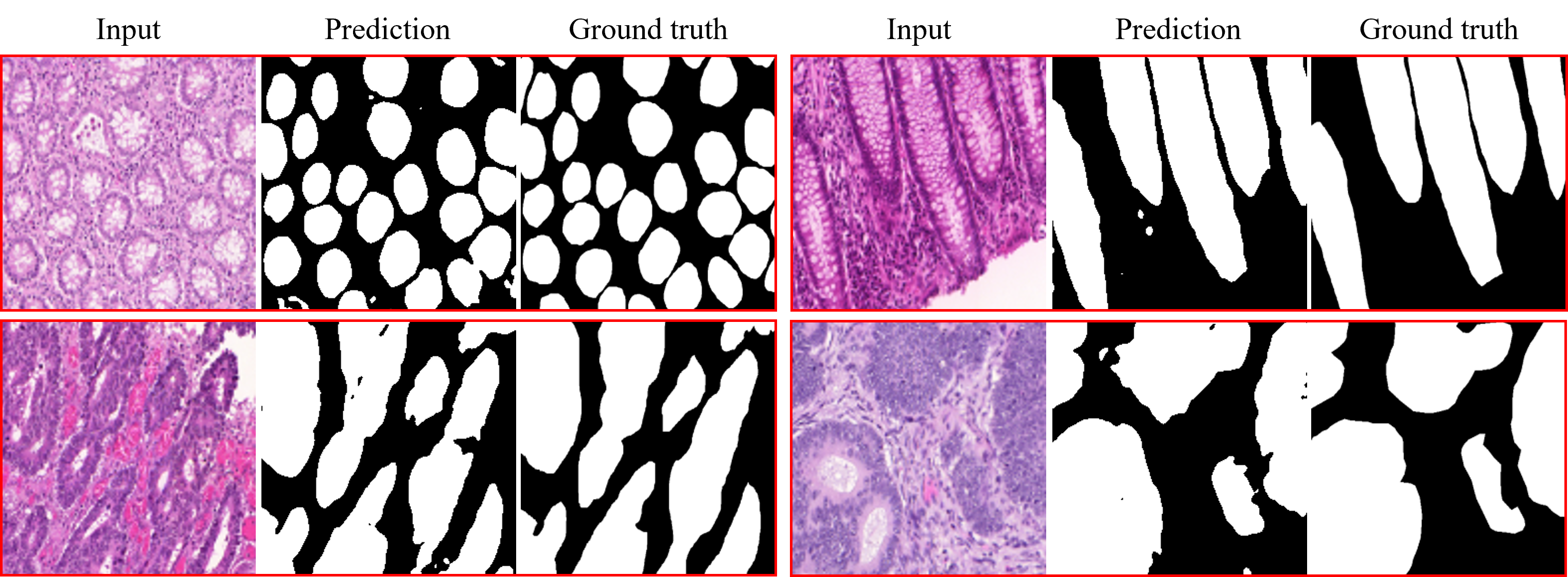}
    \caption{Visualization of predictions produced by the seUNet-Trans-M on the GlaS dataset.}
    \label{fig.subfig_Glasextra}
  \end{subfigure}
  
  \begin{subfigure}{\textwidth}
    \centering
    \includegraphics[width=0.95\linewidth]{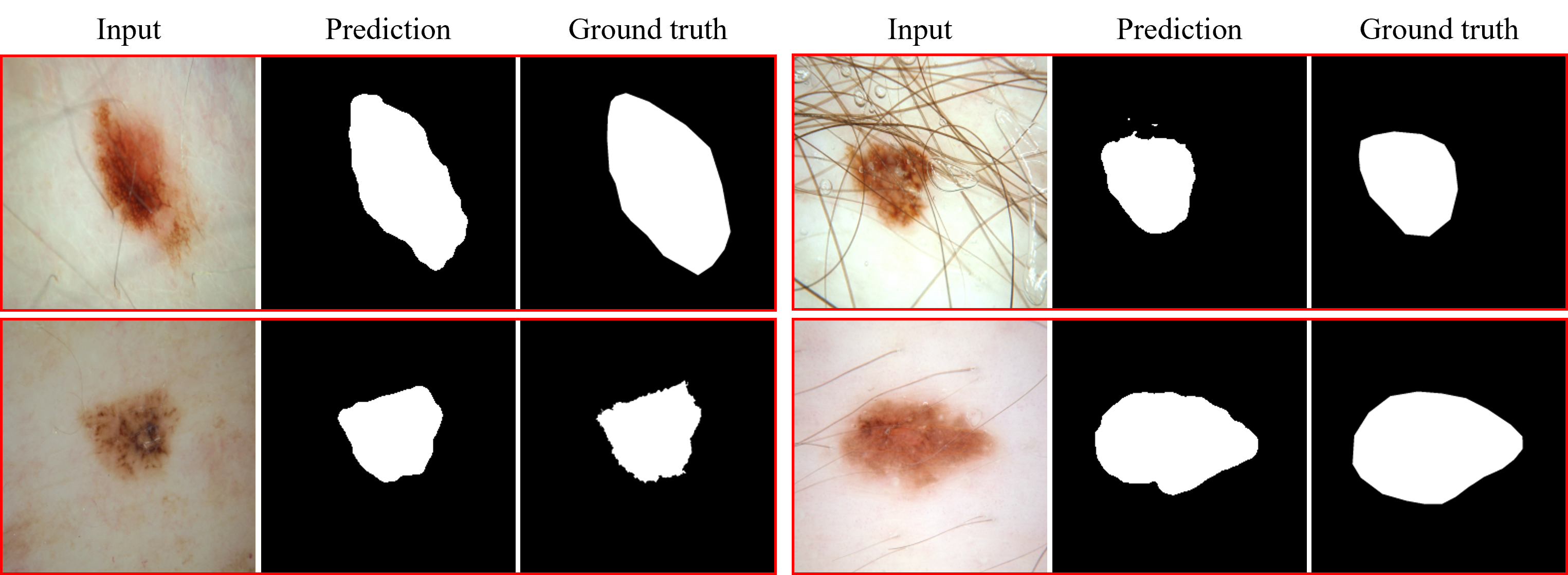}
    \caption{Visualization of predictions produced by the seUNet-Trans-M on the ISIC2018 dataset.}
    \label{fig.subfig_ISIC2018extra}
  \end{subfigure}
  
  \begin{subfigure}{\textwidth}
    \centering
    \includegraphics[width=0.95\linewidth]{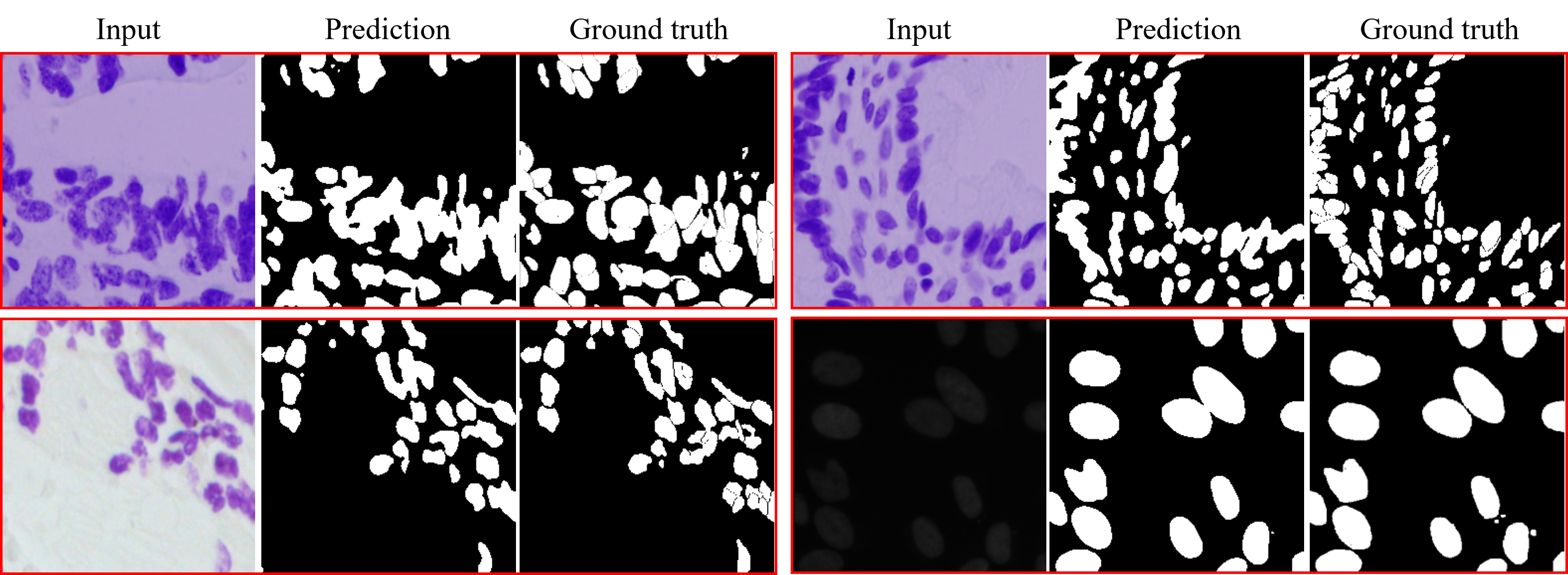}
    \caption{Visualization of predictions produced by the seUNet-Trans-M on the 2018 Data Science Bowl dataset.}
    \label{fig.subfig_2018Bowlextra}
  \end{subfigure}
  
  \caption{Qualitative comparison of different predictions on different datasets by visualization. On a specific set of images covered by a red rectangular, the input, prediction, and ground are organized from the left to the right.}
  \label{fig.combined-extraresult}
\end{figure*}

Figure \ref{fig.combined-extraresult} shows the predictions of our seUNet-Trans on different datasets. Even when dealing with relatively small datasets such as GlaS or the 2018 Data Science Bowl, the model illustrates impressive performance in comparison to the ground truth. Notably, in the case of the 2018 Data Science Bowl, as depicted in Figure \ref{fig.subfig_2018Bowlextra}, our model shows the ability to recognize mislabeling, even when the ground truth does not precisely align with the input image.

Similarly, as illustrated in Figure \ref{fig.subfig_ISIC2018extra}, our model also demonstrates its capability on the prediction of the ISIC2018 dataset. These predictions, generated by the seUNet-Trans-M, closely adhere to the object boundaries within the input images, rather than rigidly following the ground truth. These results indicate the model's proficiency in providing precise predictions across various datasets, thereby promoting its potential as a strong tool for medical image segmentation.

\section{Conclusion and Discussion}
\label{sec.6}
This paper introduces an innovative approach, named as seUNet-Trans,  that synergizes the robust feature extraction capabilities of CNN with the sophisticated contextual understanding of Transformer-based models to advance medical image segmentation. Our seUNet-Trans employs a hybrid design, integrating a fully convolutional network, UNet, with a Transformer-based model. At the heart of this integration is a specially designed bridge layer that acts as a bridge, sequentially channeling rich feature maps from UNet into the Transformer. This design enables the framework to leverage the spatial hierarchies recognized by UNet and the global dependencies discerned by the Transformer, providing a more precise and context-aware segmentation performance.

In our approach, we streamline the architecture of the proposed model by adopting a pixel-level embedding technique that forgoes the traditional use of position embeddings. Furthermore, we explore the trade-off between computational complexity and model accuracy by employing a computational reduction technique, resulting in the creation of three distinct models (L, M, and S). Such a design can enhance the model's efficiency, as it reduces the complexity of the input representation while maintaining the inherent spatial relationships of the pixels. This simplification is predicated on the understanding that, in the context of medical image segmentation, the relative positioning of pixel data is often implicit within the pixel intensity and texture patterns, making separate positional encoding redundant. As a result, our model remains attuned to the crucial spatial cues necessary for accurate segmentation without the computational overhead typically introduced by position embeddings and the original Transformer.

We have rigorously evaluated the performance of seUNet-Trans through comprehensive experiments spanning seven diverse datasets. The numerical outcomes from these experiments clearly demonstrate that our proposed model not only meets but also surpasses the benchmarks set by other state-of-the-art models in a majority of the tests. This is particularly notable in mixed Polyp segmentation tasks, where the seUNet-Trans model exhibits superior proficiency, underscoring its robustness and effectiveness in handling complex image segmentation challenges.

The promising results presented in this study pave the way for the application of our proposed model across a wider range of tasks. Future endeavors will focus on the development of specialized, lightweight versions of seUNet-Trans for specific application needs. Moreover, we will investigate the integration of advanced techniques such as the Swin Transformer, which holds the potential to elevate the efficacy of our model even further. Such explorations are expected to yield significant contributions to the field of medical image analysis and beyond.

\bibliographystyle{IEEEtran}  
\bibliography{Cite}

\begin{thebibliography}{10}
\providecommand{\url}[1]{#1}
\csname url@samestyle\endcsname
\providecommand{\newblock}{\relax}
\providecommand{\bibinfo}[2]{#2}
\providecommand{\BIBentrySTDinterwordspacing}{\spaceskip=0pt\relax}
\providecommand{\BIBentryALTinterwordstretchfactor}{4}
\providecommand{\BIBentryALTinterwordspacing}{\spaceskip=\fontdimen2\font plus
\BIBentryALTinterwordstretchfactor\fontdimen3\font minus \fontdimen4\font\relax}
\providecommand{\BIBforeignlanguage}[2]{{%
\expandafter\ifx\csname l@#1\endcsname\relax
\typeout{** WARNING: IEEEtran.bst: No hyphenation pattern has been}%
\typeout{** loaded for the language `#1'. Using the pattern for}%
\typeout{** the default language instead.}%
\else
\language=\csname l@#1\endcsname
\fi
#2}}
\providecommand{\BIBdecl}{\relax}
\BIBdecl

\bibitem{declinically}
J.~De~Fauw \emph{et~al.}, ``Clinically applicable deep learning for diagnosis and referral in retinal disease, august 2018.''

\bibitem{ouyang2020video}
D.~Ouyang, B.~He, A.~Ghorbani, N.~Yuan, J.~Ebinger, C.~P. Langlotz, P.~A. Heidenreich, R.~A. Harrington, D.~H. Liang, E.~A. Ashley \emph{et~al.}, ``Video-based ai for beat-to-beat assessment of cardiac function,'' \emph{Nature}, vol. 580, no. 7802, pp. 252--256, 2020.

\bibitem{hesamian2019deep}
M.~H. Hesamian, W.~Jia, X.~He, and P.~Kennedy, ``Deep learning techniques for medical image segmentation: achievements and challenges,'' \emph{Journal of digital imaging}, vol.~32, pp. 582--596, 2019.

\bibitem{isensee2019automated}
F.~Isensee, P.~F. J{\"a}ger, S.~A. Kohl, J.~Petersen, and K.~H. Maier-Hein, ``Automated design of deep learning methods for biomedical image segmentation,'' \emph{arXiv preprint arXiv:1904.08128}, 2019.

\bibitem{ronneberger2015u}
O.~Ronneberger, P.~Fischer, and T.~Brox, ``U-net: Convolutional networks for biomedical image segmentation,'' in \emph{Medical Image Computing and Computer-Assisted Intervention--MICCAI 2015: 18th International Conference, Munich, Germany, October 5-9, 2015, Proceedings, Part III 18}.\hskip 1em plus 0.5em minus 0.4em\relax Springer, 2015, pp. 234--241.

\bibitem{long2015fully}
J.~Long, E.~Shelhamer, and T.~Darrell, ``Fully convolutional networks for semantic segmentation,'' in \emph{Proceedings of the IEEE conference on computer vision and pattern recognition}, 2015, pp. 3431--3440.

\bibitem{schlemper2019attention}
J.~Schlemper, O.~Oktay, M.~Schaap, M.~Heinrich, B.~Kainz, B.~Glocker, and D.~Rueckert, ``Attention gated networks: Learning to leverage salient regions in medical images,'' \emph{Medical image analysis}, vol.~53, pp. 197--207, 2019.

\bibitem{wang2018non}
X.~Wang, R.~Girshick, A.~Gupta, and K.~He, ``Non-local neural networks,'' in \emph{Proceedings of the IEEE conference on computer vision and pattern recognition}, 2018, pp. 7794--7803.

\bibitem{vaswani2017attention}
A.~Vaswani, N.~Shazeer, N.~Parmar, J.~Uszkoreit, L.~Jones, A.~N. Gomez, {\L}.~Kaiser, and I.~Polosukhin, ``Attention is all you need,'' \emph{Advances in neural information processing systems}, vol.~30, 2017.

\bibitem{dosovitskiy2020image}
A.~Dosovitskiy, L.~Beyer, A.~Kolesnikov, D.~Weissenborn, X.~Zhai, T.~Unterthiner, M.~Dehghani, M.~Minderer, G.~Heigold, S.~Gelly \emph{et~al.}, ``An image is worth 16x16 words: Transformers for image recognition at scale,'' \emph{arXiv preprint arXiv:2010.11929}, 2020.

\bibitem{krizhevsky2012imagenet}
A.~Krizhevsky, I.~Sutskever, and G.~E. Hinton, ``Imagenet classification with deep convolutional neural networks,'' \emph{Advances in neural information processing systems}, vol.~25, 2012.

\bibitem{carion2020end}
N.~Carion, F.~Massa, G.~Synnaeve, N.~Usunier, A.~Kirillov, and S.~Zagoruyko, ``End-to-end object detection with transformers,'' in \emph{European conference on computer vision}.\hskip 1em plus 0.5em minus 0.4em\relax Springer, 2020, pp. 213--229.

\bibitem{zheng2021rethinking}
S.~Zheng, J.~Lu, H.~Zhao, X.~Zhu, Z.~Luo, Y.~Wang, Y.~Fu, J.~Feng, T.~Xiang, P.~H. Torr \emph{et~al.}, ``Rethinking semantic segmentation from a sequence-to-sequence perspective with transformers,'' in \emph{Proceedings of the IEEE/CVF conference on computer vision and pattern recognition}, 2021, pp. 6881--6890.

\bibitem{chen2021transUNet}
J.~Chen, Y.~Lu, Q.~Yu, X.~Luo, E.~Adeli, Y.~Wang, L.~Lu, A.~L. Yuille, and Y.~Zhou, ``Transunet: Transformers make strong encoders for medical image segmentation,'' \emph{arXiv preprint arXiv:2102.04306}, 2021.

\bibitem{zhang2021transfuse}
Y.~Zhang, H.~Liu, and Q.~Hu, ``Transfuse: Fusing transformers and cnns for medical image segmentation,'' in \emph{Medical Image Computing and Computer Assisted Intervention--MICCAI 2021: 24th International Conference, Strasbourg, France, September 27--October 1, 2021, Proceedings, Part I 24}.\hskip 1em plus 0.5em minus 0.4em\relax Springer, 2021, pp. 14--24.

\bibitem{zhou2018UNet++}
Z.~Zhou, M.~M. Rahman~Siddiquee, N.~Tajbakhsh, and J.~Liang, ``Unet++: A nested u-net architecture for medical image segmentation,'' in \emph{Deep Learning in Medical Image Analysis and Multimodal Learning for Clinical Decision Support: 4th International Workshop, DLMIA 2018, and 8th International Workshop, ML-CDS 2018, Held in Conjunction with MICCAI 2018, Granada, Spain, September 20, 2018, Proceedings 4}.\hskip 1em plus 0.5em minus 0.4em\relax Springer, 2018, pp. 3--11.

\bibitem{oktay2018attention}
O.~Oktay, J.~Schlemper, L.~L. Folgoc, M.~Lee, M.~Heinrich, K.~Misawa, K.~Mori, S.~McDonagh, N.~Y. Hammerla, B.~Kainz \emph{et~al.}, ``Attention u-net: Learning where to look for the pancreas,'' \emph{arXiv preprint arXiv:1804.03999}, 2018.

\bibitem{diakogiannis2020resUNet}
F.~I. Diakogiannis, F.~Waldner, P.~Caccetta, and C.~Wu, ``Resunet-a: A deep learning framework for semantic segmentation of remotely sensed data,'' \emph{ISPRS Journal of Photogrammetry and Remote Sensing}, vol. 162, pp. 94--114, 2020.

\bibitem{he2016deep}
K.~He, X.~Zhang, S.~Ren, and J.~Sun, ``Deep residual learning for image recognition,'' in \emph{Proceedings of the IEEE conference on computer vision and pattern recognition}, 2016, pp. 770--778.

\bibitem{fan2020pranet}
D.-P. Fan, G.-P. Ji, T.~Zhou, G.~Chen, H.~Fu, J.~Shen, and L.~Shao, ``Pranet: Parallel reverse attention network for polyp segmentation,'' in \emph{International conference on medical image computing and computer-assisted intervention}.\hskip 1em plus 0.5em minus 0.4em\relax Springer, 2020, pp. 263--273.

\bibitem{valanarasu2020kiu}
J.~M.~J. Valanarasu, V.~A. Sindagi, I.~Hacihaliloglu, and V.~M. Patel, ``Kiu-net: Towards accurate segmentation of biomedical images using over-complete representations,'' in \emph{Medical Image Computing and Computer Assisted Intervention--MICCAI 2020: 23rd International Conference, Lima, Peru, October 4--8, 2020, Proceedings, Part IV 23}.\hskip 1em plus 0.5em minus 0.4em\relax Springer, 2020, pp. 363--373.

\bibitem{jha2020doubleu}
D.~Jha, M.~A. Riegler, D.~Johansen, P.~Halvorsen, and H.~D. Johansen, ``Doubleu-net: A deep convolutional neural network for medical image segmentation,'' in \emph{2020 IEEE 33rd International symposium on computer-based medical systems (CBMS)}.\hskip 1em plus 0.5em minus 0.4em\relax IEEE, 2020, pp. 558--564.

\bibitem{tomar2022fanet}
N.~K. Tomar, D.~Jha, M.~A. Riegler, H.~D. Johansen, D.~Johansen, J.~Rittscher, P.~Halvorsen, and S.~Ali, ``Fanet: A feedback attention network for improved biomedical image segmentation,'' \emph{IEEE Transactions on Neural Networks and Learning Systems}, 2022.

\bibitem{strudel2021segmenter}
R.~Strudel, R.~Garcia, I.~Laptev, and C.~Schmid, ``Segmenter: Transformer for semantic segmentation,'' in \emph{Proceedings of the IEEE/CVF international conference on computer vision}, 2021, pp. 7262--7272.

\bibitem{xie2021segformer}
E.~Xie, W.~Wang, Z.~Yu, A.~Anandkumar, J.~M. Alvarez, and P.~Luo, ``Segformer: Simple and efficient design for semantic segmentation with transformers,'' \emph{Advances in Neural Information Processing Systems}, vol.~34, pp. 12\,077--12\,090, 2021.

\bibitem{valanarasu2021medical}
J.~M.~J. Valanarasu, P.~Oza, I.~Hacihaliloglu, and V.~M. Patel, ``Medical transformer: Gated axial-attention for medical image segmentation,'' in \emph{Medical Image Computing and Computer Assisted Intervention--MICCAI 2021: 24th International Conference, Strasbourg, France, September 27--October 1, 2021, Proceedings, Part I 24}.\hskip 1em plus 0.5em minus 0.4em\relax Springer, 2021, pp. 36--46.

\bibitem{liu2021swin}
Z.~Liu, Y.~Lin, Y.~Cao, H.~Hu, Y.~Wei, Z.~Zhang, S.~Lin, and B.~Guo, ``Swin transformer: Hierarchical vision transformer using shifted windows,'' in \emph{Proceedings of the IEEE/CVF international conference on computer vision}, 2021, pp. 10\,012--10\,022.

\bibitem{lin2022ds}
A.~Lin, B.~Chen, J.~Xu, Z.~Zhang, G.~Lu, and D.~Zhang, ``Ds-transunet: Dual swin transformer u-net for medical image segmentation,'' \emph{IEEE Transactions on Instrumentation and Measurement}, vol.~71, pp. 1--15, 2022.

\bibitem{lecun1995convolutional}
Y.~LeCun, Y.~Bengio \emph{et~al.}, ``Convolutional networks for images, speech, and time series,'' \emph{The handbook of brain theory and neural networks}, vol. 3361, no.~10, p. 1995, 1995.

\bibitem{ioffe2015batch}
S.~Ioffe and C.~Szegedy, ``Batch normalization: Accelerating deep network training by reducing internal covariate shift,'' in \emph{International conference on machine learning}.\hskip 1em plus 0.5em minus 0.4em\relax pmlr, 2015, pp. 448--456.

\bibitem{nair2010rectified}
V.~Nair and G.~E. Hinton, ``Rectified linear units improve restricted boltzmann machines,'' in \emph{Proceedings of the 27th international conference on machine learning (ICML-10)}, 2010, pp. 807--814.

\bibitem{pradhyumna2021graph}
P.~Pradhyumna, G.~Shreya \emph{et~al.}, ``Graph neural network (gnn) in image and video understanding using deep learning for computer vision applications,'' in \emph{2021 Second International Conference on Electronics and Sustainable Communication Systems (ICESC)}.\hskip 1em plus 0.5em minus 0.4em\relax IEEE, 2021, pp. 1183--1189.

\bibitem{wang2021pyramid}
W.~Wang, E.~Xie, X.~Li, D.-P. Fan, K.~Song, D.~Liang, T.~Lu, P.~Luo, and L.~Shao, ``Pyramid vision transformer: A versatile backbone for dense prediction without convolutions,'' in \emph{Proceedings of the IEEE/CVF international conference on computer vision}, 2021, pp. 568--578.

\bibitem{hendrycks2016gaussian}
D.~Hendrycks and K.~Gimpel, ``Gaussian error linear units (gelus),'' \emph{arXiv preprint arXiv:1606.08415}, 2016.

\bibitem{jha2021real}
D.~Jha, S.~Ali, N.~K. Tomar, H.~D. Johansen, D.~Johansen, J.~Rittscher, M.~A. Riegler, and P.~Halvorsen, ``Real-time polyp detection, localization and segmentation in colonoscopy using deep learning,'' \emph{Ieee Access}, vol.~9, pp. 40\,496--40\,510, 2021.

\bibitem{kingma2014adam}
D.~P. Kingma and J.~Ba, ``Adam: A method for stochastic optimization,'' \emph{arXiv preprint arXiv:1412.6980}, 2014.

\bibitem{xiao2018weighted}
X.~Xiao, S.~Lian, Z.~Luo, and S.~Li, ``Weighted res-unet for high-quality retina vessel segmentation,'' in \emph{2018 9th international conference on information technology in medicine and education (ITME)}.\hskip 1em plus 0.5em minus 0.4em\relax IEEE, 2018, pp. 327--331.

\bibitem{jha2019resUNet++}
D.~Jha, P.~H. Smedsrud, M.~A. Riegler, D.~Johansen, T.~De~Lange, P.~Halvorsen, and H.~D. Johansen, ``Resunet++: An advanced architecture for medical image segmentation,'' in \emph{2019 IEEE international symposium on multimedia (ISM)}.\hskip 1em plus 0.5em minus 0.4em\relax IEEE, 2019, pp. 225--2255.

\bibitem{zhao2017pyramid}
H.~Zhao, J.~Shi, X.~Qi, X.~Wang, and J.~Jia, ``Pyramid scene parsing network,'' in \emph{Proceedings of the IEEE conference on computer vision and pattern recognition}, 2017, pp. 2881--2890.

\bibitem{wang2020deep}
J.~Wang, K.~Sun, T.~Cheng, B.~Jiang, C.~Deng, Y.~Zhao, D.~Liu, Y.~Mu, M.~Tan, X.~Wang \emph{et~al.}, ``Deep high-resolution representation learning for visual recognition,'' \emph{IEEE transactions on pattern analysis and machine intelligence}, vol.~43, no.~10, pp. 3349--3364, 2020.

\bibitem{chen2018encoder}
L.-C. Chen, Y.~Zhu, G.~Papandreou, F.~Schroff, and H.~Adam, ``Encoder-decoder with atrous separable convolution for semantic image segmentation,'' in \emph{Proceedings of the European conference on computer vision (ECCV)}, 2018, pp. 801--818.

\bibitem{huang2021hardnet}
C.-H. Huang, H.-Y. Wu, and Y.-L. Lin, ``Hardnet-mseg: A simple encoder-decoder polyp segmentation neural network that achieves over 0.9 mean dice and 86 fps,'' \emph{arXiv preprint arXiv:2101.07172}, 2021.

\bibitem{fang2019selective}
Y.~Fang, C.~Chen, Y.~Yuan, and K.-y. Tong, ``Selective feature aggregation network with area-boundary constraints for polyp segmentation,'' in \emph{Medical Image Computing and Computer Assisted Intervention--MICCAI 2019: 22nd International Conference, Shenzhen, China, October 13--17, 2019, Proceedings, Part I 22}.\hskip 1em plus 0.5em minus 0.4em\relax Springer, 2019, pp. 302--310.

\bibitem{zhang2018road}
Z.~Zhang, Q.~Liu, and Y.~Wang, ``Road extraction by deep residual u-net,'' \emph{IEEE Geoscience and Remote Sensing Letters}, vol.~15, no.~5, pp. 749--753, 2018.

\bibitem{badrinarayanan2017segnet}
V.~Badrinarayanan, A.~Kendall, and R.~Cipolla, ``Segnet: A deep convolutional encoder-decoder architecture for image segmentation,'' \emph{IEEE transactions on pattern analysis and machine intelligence}, vol.~39, no.~12, pp. 2481--2495, 2017.

\bibitem{oktay1804attention}
O.~Oktay, J.~Schlemper, L.~L. Folgoc, M.~Lee, M.~Heinrich, K.~Misawa, K.~Mori, S.~McDonagh, N.~Y. Hammerla, B.~Kainz \emph{et~al.}, ``Attention u-net: Learning where to look for the pancreas. arxiv 2018,'' \emph{arXiv preprint arXiv:1804.03999}, 1804.

\bibitem{alom2018recurrent}
M.~Z. Alom, M.~Hasan, C.~Yakopcic, T.~M. Taha, and V.~K. Asari, ``Recurrent residual convolutional neural network based on u-net (r2u-net) for medical image segmentation,'' \emph{arXiv preprint arXiv:1802.06955}, 2018.

\bibitem{azad2019bi}
R.~Azad, M.~Asadi-Aghbolaghi, M.~Fathy, and S.~Escalera, ``Bi-directional convlstm u-net with densley connected convolutions,'' in \emph{Proceedings of the IEEE/CVF international conference on computer vision workshops}, 2019, pp. 0--0.

\end{thebibliography}

\end{document}